\documentclass[prd,preprint,nofootinbib]{revtex4-1}

\usepackage[normalem]{ulem}
\usepackage{amssymb,amsmath}
\usepackage{color}
\usepackage{graphicx,subfigure}
\usepackage{hyperref}
\usepackage{float} 
\usepackage[T1]{fontenc}
\usepackage{multirow}

\newcommand{\nn}{\nonumber}

\newcommand \ket[1]{
        \left| #1 \right>
}

\newcommand{\Dbar}{\,\overline{\!D}}

\newcommand{\DorDbar}{\raisebox{7.7pt}{$\scriptscriptstyle(\hspace*{8.5pt})$}
  \hspace*{-10.7pt}\!\Dbar}

\newcommand{\beq}{\begin{eqnarray}}
\newcommand{\eeq}{\end{eqnarray}}
\newcommand{\be}{\begin{eqnarray}}
\newcommand{\ee}{\end{eqnarray}}
\def\beqa{\begin{eqnarray}}
\def\eeqa{\end{eqnarray}}
\def\bea{\begin{eqnarray}}
\def\eea{\end{eqnarray}}

\newcommand{\bv}{\left(\begin{array}{c}}
\newcommand{\ev}{\end{array}\right)}
\newcommand{\bmtwo}{\left(\begin{array}{cc}}
\newcommand{\bmthree}{\left(\begin{array}{ccc}}
\newcommand{\emn}{\end{array}\right)}
\newcommand{\bmtwoc}{\left\{\begin{array}{cc}}
\newcommand{\bmthreec}{\left\{\begin{array}{ccc}}
\newcommand{\emnc}{\end{array}\right\}}
\newcommand{\ba}{\begin{array}}
\newcommand{\ea}{\end{array}}

\begin{document}

\title{Probing the $\Delta U=0$ Rule in Three Body Charm Decays}

\preprint{MAN/HEP/2020/015}

\author{Avital Dery}
\email{avital.dery@cornell.edu}
\affiliation{Department of Physics, LEPP, Cornell University, Ithaca, NY 14853, USA}

\author{Yuval Grossman}
\email{yg73@cornell.edu}
\affiliation{Department of Physics, LEPP, Cornell University, Ithaca, NY 14853, USA}

\author{Stefan Schacht}
\email{stefan.schacht@manchester.ac.uk}
\affiliation{Department of Physics and Astronomy, University of Manchester, Manchester M13 9PL, United Kingdom}

\author{Abner Soffer}
\email{asoffer@tau.ac.il}
\affiliation{School of Physics and Astronomy, Tel Aviv University, Tel Aviv 69978, Israel}

\begin{abstract}
CP violation in charm decay was observed in the decays $D^0\rightarrow P^{\pm}P^{\mp}$ of a $D^0$ meson to two pseudoscalars. When interpreted within the SM, the results imply that the ratio of the relevant rescattering amplitudes has a magnitude and phase that are both of $O(1)$. We discuss ways to probe similar ratios in $D^0\rightarrow V^{\pm}P^{\mp}$ decays, where $V$ is a vector that decays to two pseudoscalars, from the Dalitz-plot analysis of time-integrated three-body decays. Compared to two-body decays, three-body decays have the advantage that the complete system can be solved without the need for time-dependent CP violation measurements or use of correlated $D^0$--$\overline{D}^0$ production. We discuss the decays $D^0\rightarrow \pi^+\pi^-\pi^0$ and  $D^0\rightarrow K^+K^-\pi^0$ as examples by considering a toy model of only two overlapping charged resonances, treating the underlying pseudo two-body decays in full generality.
\end{abstract}

\maketitle

\section{Introduction \label{sec:intro}}

The discovery of charm CP violation~\cite{Aaij:2019kcg} in singly-Cabibbo suppressed (SCS) two-body $D\rightarrow PP$ decays, where $P$ is a 
pseudoscalar meson, was the result of a  
multi-year effort~\cite{Aitala:1997ff, Link:2000aw, Csorna:2001ww, Aubert:2007if, Staric:2008rx, Aaltonen:2011se, Collaboration:2012qw, Aaij:2011in, Aaij:2013bra, Aaij:2014gsa, Aaij:2016cfh, Aaij:2016dfb} and triggered immediate theoretical 
interpretations~\cite{Grossman:2019xcj,Chala:2019fdb, Li:2019hho,Soni:2019xko}. Works before the discovery can be found in Refs.~\cite{Einhorn:1975fw, Abbott:1979fw,Golden:1989qx,  Brod:2012ud, Grinstein:2014aza, Bhattacharya:2012ah, Franco:2012ck, Hiller:2012xm, Nierste:2017cua, Nierste:2015zra, Muller:2015rna, Grossman:2018ptn, Buccella:1994nf, Grossman:2006jg, Artuso:2008vf, Khodjamirian:2017zdu, Buccella:2013tya, Cheng:2012wr, Feldmann:2012js, Li:2012cfa, Atwood:2012ac, Grossman:2012ry, Buccella:2019kpn, Yu:2017oky, Brod:2011re, Pirtskhalava:2011va}. 
In the language of the approximate SU(3)$_F$ symmetry of QCD and its SU(2)-subgroup of 
unitary rotations of the $s$ and $d$ quarks, called $U$-spin, the new measurement enables the extraction of the ratio of  
$\Delta U=0$ over $\Delta U=1$ hadronic matrix elements. 
Parametrizing this ratio as 
\begin{align}
\widetilde{r}_0 &= 1 + C e^{i\delta},
\end{align}
with $C$ a real number and $\delta$ a strong phase, we can differentiate between three scenarios~\cite{Grossman:2019xcj}:
\begin{align}
1)\,\, C=\mathcal{O}(\alpha_s/\pi)\,, \qquad
2)\,\, C=\mathcal{O}(1)\,, \qquad
3)\,\, C\gg \mathcal{O}(1)\,.
\end{align}
For a generic phase of magnitude $\delta\sim \mathcal{O}(1)$, the observed result for CP violation in two-body decays~\cite{Aaij:2019kcg} is consistent with option (2), \emph{i.e.}~$C\sim 1$. 
Note that options (2) and (3) are not different on a principle basis, as both describe non-perturbative physics. 
Note further, that a $C$-value of category~(1) would correspond to $\Delta a_{CP}^{\mathrm{dir}} \sim 1\cdot 10^{-4}$, in contrast to the measurement of $\Delta a_{CP}^{\mathrm{dir}} = -0.00164\pm 0.00028$~\cite{Aaij:2019kcg, Amhis:2019ckw}.
This means that if we have a strong argument for $C$ being of category (1) the current measurement would imply new physics~\cite{Grossman:2019xcj}.

Option (2) can be interpreted as $KK\leftrightarrow \pi\pi$ rescattering, with potential contributions from all on-shell multi-body light hadronic states that can rescatter into $KK$ and $\pi\pi$.
This large rescattering was named ``the
$\Delta U=0$ rule'' in Refs.~\cite{Brod:2012ud, Grossman:2012ry,
  Grossman:2019xcj}, in analogy to the well-known $\Delta I=1/2$ rule
in kaon decays. The latter corresponds to
large non-trivial rescattering effects and belongs to category (3)
above~\cite{Tanabashi:2018oca, GellMann:1955jx, GellMann:1957wh,
  Gaillard:1974nj, Bardeen:1986vz, Buras:2014maa, Bai:2015nea,
  Blum:2015ywa,Boyle:2012ys, Buras:2015yba, Kitahara:2016nld,
  Abbott:2020hxn}. Note that while the underlying dynamical mechanisms
of the $\Delta U=0$ and $\Delta I=1/2$ rule are not identical, because of
the different scales entering the processes, the main conclusion is
the same: rescattering is important, and naive quark level calculations
are not adequate in such cases.

The interpretation of the data as being of category (2) relies on the assumption of a 
generic $\mathcal{O}(1)$ strong phase between the 
$\Delta U=0$ and $\Delta U=1$ matrix elements. Yet, as of now we do
not have any information about that strong phase, which can only be determined in future 
time-dependent measurements or in correlated $D^0$--$\overline{D}^0$ decays, see Ref.~\cite{Grossman:2019xcj} for a 
detailed discussion.

Given that the charm quark is not very heavy relative to the scale of QCD, one expects rescattering in nonleptonic charm decays to be indeed of $\mathcal{O}(1)$. 
However, there is no hard argument from 
first principles that this is necessarily the case.
For now, it is an open question whether the observed effect is a sign of new physics or not~\cite{Grossman:2019xcj,Chala:2019fdb,Khodjamirian:2017zdu}. 
It is hence very important to extract the analogous ratios of $\Delta U=0$ over $\Delta U=1$ hadronic matrix elements
in additional sets of decays, such as $D^0\rightarrow V^{\pm}P^{\mp}$, where $V$ is a vector meson.
In fact, $D^0\rightarrow V^{\pm}P^{\mp}$ decays have an important advantage over $D^0\rightarrow P^{\pm}P^{\mp}$:  as pointed out long ago in Ref.~\cite{Snyder:1993mx} and discussed also in  
Refs.~\cite{Lipkin:1991st, Bigi:2015gca, Charles:2017ptc, Bediaga:2009tr, Nogueira:2015tsa, Fajfer:2004cx, Atwood:2012ac}, the interference of two-body decays leading to a common three-body final state can be used to extract strong phases and penguin amplitudes. 
Throughout, we refer to $D^0\rightarrow P^{\pm}P^{\mp}$ decays as \lq\lq{}two-body\rq\rq{} decays, and to $D^0\rightarrow V^{\pm}P^{\mp}$ decays as \lq\lq{}pseudo two-body\rq\rq{} decays.

In this article we present a method for measuring the effects of rescattering in the charm-decay Dalitz plot and 
investigating the $\Delta U=0$ rule in $D^0\rightarrow V^{\pm}P^{\mp}$ decays. Most importantly, the interference effects enable the determination of the relative phase between the $\Delta U=0$ and $\Delta U=1$ matrix elements even with time-integrated CP violation measurements. 
Therefore, contrary to the case of $D^0\rightarrow P^+P^-$ decays, no time-dependent measurements are required for this purpose.
As examples, we focus here on the decays 
$D^0\rightarrow \pi^+\pi^-\pi^0$~\cite{CroninHennessy:2005sy,Aubert:2006xn,Aubert:2007ii,Arinstein:2008zh, Aubert:2008yd, Nayak:2014tea,Malde:2015mha, Aaij:2014afa, Aaij:2015jna} 
and  
$D^0\rightarrow K^+K^-\pi^0$~\cite{Aubert:2007dc,  Aubert:2008yd, Aubert:2006xn, Cawlfield:2006hm, Asner:1996ht, Nayak:2014tea, Malde:2015mha, Aaij:2014afa, Aaij:2015jna}. We employ a toy model using only two overlapping resonances and treat $D^0\rightarrow V^{\pm}P^{\mp}$ as pseudo two-body decays, highlighting the insights that can be gained from the interference effects. 
Future work may treat the Dalitz plot in a more complete way, especially including also neutral resonances. 

Note that our approach treats the pseudo two-body decays in a model-independent way, and our theory assumptions concern mainly how to model the three-body decays in terms of the pseudo two-body decays. For example, because of that, we do not include rescattering data of $KK\leftrightarrow \pi\pi$, which is available,~\emph{e.g.}~in Ref.~\cite{Pelaez:2016tgi}. Using this data for charm decays is only possible with further theory assumptions on the factorization of charm decays, which is challenging due to the intermediate mass scale of the charm quark. 
Note further, that for such an analysis, rescattering data close to $m_{D^0} = 1.86483\pm 0.00005$ GeV~\cite{PDG2020} is needed.  

Previous isospin and SU(3)$_F$ analyses of $D\rightarrow PPP$ and $D\rightarrow PV$ decays are given in Refs.~\cite{Grossman:2012ry, Gaspero:2008rs, Grossman:2012eb, Grossman:2012ry, Bhattacharya:2010id, Bhattacharya:2010ji, Bhattacharya:2010tg, Bhattacharya:2008ke,Cheng:2010ry,Cheng:2019ggx}. 
In Ref.~\cite{Rosner:2003yk} the measurement of the relative strong phase between $D^0\rightarrow K^{*+}K^-$ and $D^0\rightarrow K^{*-}K^+$ decays is discussed. 
CP violation in $D^0\rightarrow K^+K^-\pi^0$ decays has been discussed in Ref.~\cite{Zhou:2018suj} within the factorization-assisted topological approach. 
Three-body $B$ decays have been treated in  
Refs.~\cite{Grossman:2003qp, Gronau:2003ep, Engelhard:2005hu, Engelhard:2005ky,Bhattacharya:2013boa, Bhattacharya:2014eca, Bhattacharya:2015uua,  Klein:2017xti,  Xu:2013rua, Xu:2013dta, Cheng:2016shb, Wang:2015ula, Li:2016tpn,Wang:2016rlo,Wang:2014ira, Cheng:2013dua, He:2014xha, Sahoo:2015msa, Mannel:2020abt}. Recent LHCb results on CP violation in $B^+\rightarrow \pi^+\pi^-\pi^+$ can be found in Refs.~\cite{Aaij:2019hzr, Aaij:2019jaq}. Rescattering in three-body decays is discussed in Refs.~\cite{Aoude:2018zty, Furman:2005xp}.

The question we wish to ask is as follows: can we determine that the ratio of $\Delta U=0$ over $\Delta U=1$ matrix elements in $D\rightarrow PV$ decays is $\mathcal{O}(1)$, as was found for $D\rightarrow PP$ decays in Ref.~\cite{Grossman:2019xcj}?
The answer to this question does not require a precision measurement of the respective matrix elements. 
Consequently, our methodology applies to the first round of data analyses, with experimental errors in the range of 20\%--30\%.
For this purpose, we employ a model-dependent parametrization of three-body decays, which will have to be revised when experiments collect sufficient data for precision measurements. 

For concreteness, we use the Breit-Wigner parametrization to describe the $V$ resonance. Furthermore, we neglect subdominant contributions in the region of interest of the Dalitz plot. 
Our main results are independent of the choice of the resonance parametrization, and should be applied within a more complete Dalitz analysis to future data. Note that using multiple Breit-Wigner distributions can lead to the violation of unitarity~\cite{Tanabashi:2018oca, Hanhart:2015zyp, Daub:2015xja, Niecknig:2015ija, Niecknig:2017ylb, Magalhaes:2011sh, Magalhaes:2015fva, Boito:2017jav}. However, in practice, there are many cases in which this is not a problem~\cite{Aubert:2007ii,Aubert:2007dc}.  
In any case, the parametrization that we employ for the underlying pseudo two-body 
decays is completely general and is therefore valid also for future precision studies. 

After introducing our notation in Sec.~\ref{sec:notation}, 
we discuss in Sec.~\ref{sec:minimal} the complete set of observables for $D^0\rightarrow\rho^{\pm}\pi^{\mp}$ and the extraction of the $\Delta U=0$ over $\Delta U=1$ ratio of matrix elements, including their relative phase, from  $D^0\rightarrow \pi^+\pi^-\pi^0$ decays. 
In Sec.~\ref{sec:breaking} we turn to the combined Dalitz-plot analysis of $D^0\rightarrow \pi^+\pi^-\pi^0$ and $D^0\rightarrow K^+K^-\pi^0$ and the corresponding implications of $U$-spin conservation. We conclude in Sec.~\ref{sec:conclusions}.
In the appendix, we list the complete set of observables as well as parametrizations for 
$D^0\rightarrow P^+P^-$ and $D^0\rightarrow P^{\pm}V^{\mp}$ decays.

\section{Notation \label{sec:notation}}

For the purpose of naming the hadronic parameters, as well as their interpretation in terms of $U$-spin parameters, we 
use the following conventions for the U-spin doublets~\cite{Gronau:1994rj, Gronau:1995hm, Soni:2006vi}
\begin{align}
\ket{d} &= \ket{\frac{1}{2}, \frac{1}{2}}\,, & 
\ket{s} &= \ket{\frac{1}{2}, -\frac{1}{2}}\,, \\ 
 \ket{\overline{s}} &= \ket{\frac{1}{2},\frac{1}{2} }\,, & 
-\ket{\overline{d}} &= \ket{\frac{1}{2},-\frac{1}{2} }\,,
\end{align}
and following thereof\,,
\begin{align}
\ket{K^+}   &= \ket{u\overline{s}}^{P1} = \ket{\frac{1}{2}, \frac{1}{2}}^{P1}\,, &
\ket{\pi^+} &= \ket{u\overline{d}}^{P1} = -\ket{\frac{1}{2},-\frac{1}{2}}^{P1}\,, \label{eq:doublets-1}\\
\ket{\pi^-} &= -\ket{d\overline{u}}^{P2} = -\ket{\frac{1}{2},\frac{1}{2}}^{P2}\,, &
\ket{K^-}   &= -\ket{s\overline{u}}^{P2} =  \ket{\frac{1}{2},-\frac{1}{2}}^{P2}\,, \label{eq:doublets-2}\\
\ket{K^{*+}} &=  \ket{u\overline{s}}^{V1} =  \ket{\frac{1}{2}, \frac{1}{2}}^{V1}\,, &
\ket{\rho^+}  &= \ket{u\overline{d}}^{V1} = -\ket{\frac{1}{2},-\frac{1}{2}}^{V1}\,, \label{eq:doublets-3}\\
\ket{\rho^-} &= -\ket{d\overline{u}}^{V2} = -\ket{\frac{1}{2},\frac{1}{2}}^{V2}\,,  &
\ket{K^{*-}} &= -\ket{s\overline{u}}^{V2} =  \ket{\frac{1}{2},-\frac{1}{2}}^{V2}\,. \label{eq:doublets-4}
\end{align}
Note that matrix elements involving different doublets have to be distinguished and in general are not related.
For example, the charge of different doublets is different.
That is why we denote states of different doublets with extra superscripts. The same superscripts are used 
in the corresponding matrix elements below. 
Note further that the states and matrix elements of vector mesons are not related to the ones of the pseudoscalars.
For clarity, we therefore use \lq\lq{}$V$\rq\rq{} and \lq\lq{}$P$\rq\rq{} in the superscript notation, respectively. 

The Hamiltonian for singly-Cabibbo suppressed (SCS) decays is 
\begin{align}
\mathcal{H}_{\mathrm{eff}} \sim \lambda_{sd} \mathcal{H}_{(1,0)} - \frac{\lambda_b}{2} \mathcal{H}_{(0,0)}\,, \label{eq:SCS-Hamiltonian}
\end{align}
where we use the standard notation in which $\mathcal{H}_{(1,0)}$
($\mathcal{H}_{(0,0)}$) is the triplet (singlet) Hamiltonian with
$\Delta U=1$ ($\Delta U=0$) and zero third component of $U$-spin, and
we have defined the CKM matrix-element combinations
\begin{align}
\lambda_{sd} &\equiv \frac{V_{cs}^* V_{us} - V_{cd}^* V_{ud}}{2}\,, &
-\frac{\lambda_b}{2} &\equiv -\frac{V_{cb}^* V_{ub}}{2} = \frac{V_{cs}^* V_{us} + V_{cd}^* V_{ud}}{2}\,. \label{eq:CKM-combinations}
\end{align}
Note that $\lambda_{sd}\approx \lambda_{sd}^* \approx \lambda$, with the Wolfenstein parameter $\lambda$.
In the following we also use the ratio 
\begin{align}
\widetilde{\lambda}_b \equiv \frac{\lambda_b}{\lambda_{sd}} \approx 0.001 - i\, 0.003\,.
\end{align}

\section{Probing the $\Delta U=0$ rule \label{sec:minimal}}

\subsection{Overview \label{sec:counting}}

Before we discuss the investigation of the $\Delta U=0$ rule in $D^0\rightarrow \pi^+\pi^-\pi^0$, we first list the independent parameters and observables of the SCS decays $D^0\rightarrow \pi^+\pi^-$ and 
$D^0\rightarrow \rho^{\pm} \pi^{\mp}$. In order to do so, we adapt the notation used in Refs.~\cite{Grossman:2019xcj,Brod:2012ud}. This list is extended in Appendix~\ref{sec:parametrization} to include Cabibbo-favored (CF) and doubly-Cabibbo suppressed (DCS) decays. 
The identified observables correspond to a complete set of real parameters that can be extracted from the data. 
We take the SM CKM matrix elements to be known from previous measurements, \emph{i.e.}~the parameters of interest are pure QCD parameters that carry a strong phase only.

We present the independent observables in the general case and in the CP limit.
We note that our results do not depend on the CP limit. However, we use it as a useful and excellent approximation to demonstrate the simple relation between experimental observables and the parameters of interest. Specifically, extraction of tree amplitudes depends mostly on charm-flavor-averaged branching ratios, where rescattering contributions are suppressed by the ratio of CKM elements $\vert V_{ub}/V_{cb}\vert^2$ and can be safely neglected. Using the extracted tree diagrams as input, the rescattering contributions can then be obtained from the CP asymmetries.
Finally, the strong-phase difference between the rescattering and tree amplitudes can be determined from coherent $D^0$--$\overline{D}^0$ decays~\cite{Gronau:2001nr, Grossman:2019xcj}, 
time-dependent CP violation measurements~\cite{Grossman:2019xcj}, or, as we show in the following, Dalitz plot analyses. The strong phase between the tree amplitudes of different SCS decay channels can be extracted only from a Dalitz plot analysis. See Appendix~\ref{sec:parametrization} for more details.

\subsection{Case I: $D^0\rightarrow \pi^+\pi^-$}

Following the notation of Refs.~\cite{Grossman:2019xcj, Brod:2012ud}, we parametrize the amplitude for $D^0\rightarrow \pi^+\pi^-$ in full generality as 
\begin{align}
\mathcal{A}(D^0\rightarrow \pi^+\pi^-) &=  - \lambda_{sd}\, T - \lambda_b\, R \,, \label{eq:Dpipi} 
\end{align}
where 
\begin{eqnarray}
T &=& \left< \pi^+\pi^- \left| \mathcal{H}_{(1,0)} \right| D^0 \right>, \nonumber\\
R &=& \left< \pi^+\pi^- \left| \mathcal{H}_{(0,0)} \right| D^0 \right>
\label{eq:TRdef}
\end{eqnarray}
are the tree and rescattering amplitudes, respectively, and $\mathcal{H}_{(1,0)}$
($\mathcal{H}_{(0,0)}$) is the triplet (singlet) Hamiltonian with
$\Delta U=1$ ($\Delta U=0$) and zero third component of $U$-spin, see Eq.~(\ref{eq:SCS-Hamiltonian}). 
The minus sign in front of $\lambda_{sd}$ in Eq.~(\ref{eq:Dpipi}) appears because $V_{cd}^*V_{ud}$ is the relevant CKM matrix combination for the tree amplitude of $D^0\rightarrow \pi^+\pi^-$. 

As pointed out on general grounds in Ref.~\cite{Grossman:1997gd}, the 
\lq\lq{}two-term weak amplitude\rq\rq{} that we provide in Eq.~(\ref{eq:Dpipi}), is completely general, in the sense that adding a third 
term with an additional relative weak and strong phase can be absorbed by redefinitions into the existing terms. 
 
In general, for $D^0\rightarrow \pi^+\pi^-$ we have three independent observables that correspond to three real parameters:
\begin{itemize}
\item We define the CP-averaged branching ratio as 
	\begin{align}
	\mathcal{B}(D\rightarrow \pi^+\pi^-) &\equiv \frac{ \mathcal{B}(D^0\rightarrow \pi^+\pi^-) + \mathcal{B}(\overline{D}^0\rightarrow \pi^+\pi^-)}{2}\,, \label{eq:untaggedBR} 
	\end{align}
	where the notation on the LHS is such that we write \lq\lq{}$D$\rq\rq{} without bars.
	On the RHS, \lq\lq{}$D^0$\rq\rq{} and \lq\lq{}$\overline{D}^0$\rq\rq{} indicate that the flavor is tagged, usually via production in the decay $D^{*+}\to D^0 \pi^+$ or its CP-conjugate. The branching fraction $\mathcal{B}(D\rightarrow \pi^+\pi^-)$ yields $\vert T\vert^2$ in the CP-limit. 	

\item The direct CP asymmetry 
	\begin{align}
	a_{CP}^{\mathrm{dir}} &\equiv 
		\frac{\vert \mathcal{A}(D^0\rightarrow \pi^+\pi^- )\vert^2 - \vert \mathcal{A}( \overline{D}^0\rightarrow \pi^+\pi^-  )\vert^2 }
		{ \vert \mathcal{A}(D^0\rightarrow \pi^+\pi^- )\vert^2 + \vert \mathcal{A}( \overline{D}^0\rightarrow \pi^+\pi^-  )\vert^2 }\,, \label{eq:dirCP}
	\end{align}
	which yields $\vert R/T\vert\times \mathrm{arg}(R/T)$, where $\mathrm{arg}(R/T)$ is the relative strong phase.

\item The phase $\mathrm{arg}\left(\frac{\mathcal{A}(\overline{D}^0\rightarrow \pi^+\pi^-)}{\mathcal{A}(D^0\rightarrow \pi^+\pi^-)}\right)$. Determination of this quantity requires use of time-dependent measurements or a coherent $\DorDbar$ initial state~\cite{Gronau:2001nr, Grossman:2019xcj}, allowing for CP violation. This observable can be used to extract the parameter $\mathrm{arg}\left(R/T\right)$.
\end{itemize}
Either $T$ or $R$ can be chosen real, because the
overall phase of the amplitude $\mathcal{A}(D^0\rightarrow \pi^+\pi^-)$ is not physical. The relative phase between $T$ and $R$, on the other hand, is physical and can be extracted from $\mathrm{arg}\left(\frac{\mathcal{A}(\overline{D}^0\rightarrow \pi^+\pi^-)}{\mathcal{A}(D^0\rightarrow \pi^+\pi^-)}\right)$.
In the CP limit, $\mathrm{arg}(\lambda_b)=0$, and we have
\begin{align}
\Gamma(\overline{D}^0\rightarrow \pi^+\pi^-) &= \Gamma(D^0\rightarrow \pi^+\pi^-)\,, \\
\mathrm{arg}\left(\frac{\mathcal{A}(\overline{D}^0\rightarrow \pi^+\pi^-)}{\mathcal{A}(D^0\rightarrow \pi^+\pi^-)}\right) &= 0\,,
\end{align}
and thus only one independent observable, namely, the CP-averaged rate $\Gamma(D\rightarrow \pi^+\pi^-)$.
In that case, $R$ can be absorbed into $T$ by a redefinition.
The above presentation of the independent observables is also included in Table~\ref{tab:obsDPP} in Appendix~\ref{sec:appendixDPP}.

\subsection{Case II: $D^0\rightarrow \rho^{\pm}\pi^{\mp}$ and $D^0\rightarrow \pi^+\pi^-\pi^0$} 
\label{subsec:exII}

Turning to the system of $D^0\rightarrow \rho^{\pm}\pi^{\mp}$ decays, we define the corresponding CP-averaged branching ratios as 
\begin{align}
\mathcal{B}(D\rightarrow \pi^+\rho^- ) &\equiv \frac{\mathcal{B}(D^0\rightarrow \pi^+\rho^- ) + \mathcal{B}(\overline{D}^0\rightarrow \pi^-\rho^+ )}{2} \,,  \label{eq:untaggedBR-PV1}\\
\mathcal{B}(D\rightarrow \pi^-\rho^+ ) &\equiv \frac{\mathcal{B}(D^0\rightarrow \pi^-\rho^+ ) + \mathcal{B}(\overline{D}^0\rightarrow \pi^+\rho^- )}{2} \,.  \label{eq:untaggedBR-PV2}
\end{align}
We write the corresponding amplitudes in full generality as 
\begin{align}
\mathcal{A}( D^0\rightarrow \pi^+ \rho^- )  &= - \lambda_{sd}\,    T^{P_1V_2} - 
						 \lambda_b\, R^{P_1V_2} \,, \label{eq:Dpirho1} \\
\mathcal{A}(  D^0\rightarrow \pi^- \rho^+ )    &= - \lambda_{sd}\,  T^{P_2V_1}  - 
						 \lambda_b\,  R^{P_2V_1} \,, \label{eq:Dpirho2} 
\end{align}
and denote the ratio of $\Delta U=0$ to $\Delta U=1$ matrix elements as
\begin{align}
\widetilde{R}^{P_iV_j}&\equiv \frac{R^{P_iV_j}}{T^{P_iV_j}}\,. \label{eq:r-ratio}
\end{align}
This parameterization involves seven real parameters, which can be obtained from experimental measurements as follows:
\begin{itemize}
\item The two CP-averaged decay rates $\mathcal{B}( D\rightarrow \pi^+ \rho^- )$ and $\mathcal{B}( D\rightarrow \pi^- \rho^+ )$, taken in the CP limit, yield $\vert T^{P_1V_2}\vert$\,, and $\vert T^{P_2V_1}\vert$, respectively. We discuss in Sec.~\ref{sec:generalmethod} how to extract the branching fractions of pseudo two-body decays from the three-body Dalitz plot. 
\item The observable $\mathrm{arg}\left(\frac{\mathcal{A}(D^0\rightarrow \pi^+ \rho^-)}{\mathcal{A}(D^0\rightarrow \pi^- \rho^+)}\right)$, and the corresponding parameter $\mathrm{arg}\left(\frac{ T^{P_2V_1}}{ T^{P_1V_2}} \right)$, can be obtained from the Dalitz plot analysis in the CP-limit.
\item The direct CP asymmetries $a_{CP}^{\mathrm{dir}}(D^0\rightarrow \pi^+\rho^-)$ and $a_{CP}^{\mathrm{dir}}(D^0 \rightarrow \pi^- \rho^+ )$, 
allowing for CP violation, yield $\vert R^{P_1V_2}\vert\,, \vert R^{P_2V_1}\vert$ times respective strong phase factors.
\item The phases $\mathrm{arg}\left(\frac{\mathcal{A}(\overline{D}^0\rightarrow \pi^+ \rho^-)}{\mathcal{A}(D^0\rightarrow \pi^+ \rho^-)} \right)$ and $\mathrm{arg}\left(\frac{\mathcal{A}(\overline{D}^0\rightarrow \pi^- \rho^+)}{\mathcal{A}(D^0\rightarrow \pi^- \rho^+}) \right)$, as well as the corresponding phases $\mathrm{arg}\left(\frac{R^{P_1V_2}}{T^{P_1V_2}}\right)$ and $\mathrm{arg}\left(\frac{ R^{P_2V_1}}{ T^{P_2V_1} }\right)$, can be extracted from a Dalitz-plot analysis, allowing for CP violation. They can also be obtained from coherent initial-state production, time-dependent measurements, as in the case of $D^0\to \pi^+\pi^-$. 
\end{itemize}
These seven observables are also listed in Table~\ref{tab:obsDpirho}. They include two CP-averaged decay rates, two direct CP asymmetries and three phases. 
Contrary to the $D\rightarrow PP$ case, the interference needed in order to measure the phases is available not only through 
$D^0$--$\overline{D}^0$ mixing in the initial state and in the time evolution, but also from the Dalitz-plot of the three-body final state $\pi^+\pi^-\pi^0$. 
Furthermore, $D\rightarrow \pi\rho$ has not just a factor of two more observables and parameters than $D\rightarrow\pi\pi$, but also the relative phase between $T^{P_1V_2}$ and $T^{P_2V_1}$. 
Note further that apriori the parameters of $D\rightarrow \pi\rho$ and $D\rightarrow\pi\pi$ are not related numerically.
In the CP limit, relations between the observables (see Table~\ref{tab:obsDpirho}) imply that the direct CP asymmetries vanish, and there are only three independent observables and real parameters, namely
\begin{align}
\vert T^{P_1V_2}\vert\,, \quad
\vert T^{P_2V_1}\vert\,, \quad
\mathrm{arg}\left(\frac{ T^{P_2V_1}}{ T^{P_1V_2}} \right)\,.
\end{align}

\begin{table}[t]
\begin{center}
\begin{tabular}{l||c|c}
\hline
\multicolumn{2}{c|}{Observables}  & \multicolumn{1}{l}{Relations in the CP limit $\mathrm{arg}(\lambda_{b}) = 0 $}   \\\hline\hline
\rule{0pt}{1em}
\multirow{2}{3.5cm}{CP-averaged branching ratios} & $\mathcal{B}(D\rightarrow \pi^+ \rho^- )$  &  \\
& $\mathcal{B}(D\rightarrow \pi^- \rho^+ )$  &   \\\hline
\rule{0pt}{1em}
\multirow{2}{3.5cm}{Direct CP asymmetries} &
$\, a^{\rm dir}_{CP}(D\rightarrow \pi^+ \rho^- )$  &  $=0$ \\
&$\,a^{\rm dir}_{CP}(D\rightarrow \pi^- \rho^+ )$  &  $=0$ \\\hline
\multirow{3}{3.5cm}{Phases} &
$\,\arg\left(\frac{\mathcal{A}(\overline{D}^0\rightarrow \pi^+ \rho^-)}{ \mathcal{A}(D^0\rightarrow \pi^+ \rho^-)} \right)$  & $=-\arg\left(\frac{\mathcal{A}(D^0\rightarrow \pi^+\rho^-)}{\mathcal{A}(D^0\rightarrow \pi^-\rho^+)}\right)$ \\
& $\,\arg\left(\frac{\mathcal{A}(\overline{D}^0\rightarrow \pi^- \rho^+)}{ \mathcal{A}(D^0\rightarrow \pi^- \rho^+)} \right)$  & $=\arg\left(\frac{\mathcal{A}(D^0\rightarrow \pi^+\rho^-)}{\mathcal{A}(D^0\rightarrow \pi^-\rho^+)}\right)$  \\
& $\,\arg\left(\frac{\mathcal{A}(D^0\rightarrow \pi^+\rho^-)}{\mathcal{A}(D^0\rightarrow \pi^-\rho^+)}\right)$  &  \\\hline\hline
\# of independent observables\,\, &\, 7       & \, 3    \\\hline\hline
\end{tabular}
\caption{Available observables for $D\rightarrow \pi^{\pm}\rho^{\mp}$ and relations between them in the CP limit. The CP-averaged branching ratios are defined in Eqs.~(\ref{eq:untaggedBR-PV1}) and~(\ref{eq:untaggedBR-PV2})
\label{tab:obsDpirho}}
\end{center}
\end{table}

\subsubsection{General Method \label{sec:generalmethod} }

Using the parametrization of the pseudo two-body weak decays $D^0\rightarrow \rho^{\mp}\pi^{\pm}$ as given in  
Eqs.~(\ref{eq:Dpirho1}, \ref{eq:Dpirho2}), we show below how to use the $D\rightarrow \pi^+\pi^-\pi^0$ 
Dalitz plot to extract the relevant parameters, 
\begin{align}
&\vert T^{P_1V_2}\vert\,, \quad
\left\vert  \widetilde{R}^{P_1V_2}\right\vert\,,\quad
\vert T^{P_2V_1}\vert\,, \quad
\left\vert  \widetilde{R}^{P_2V_1} \right\vert\,,\nn\\
&\mathrm{arg}\left( \widetilde{R}^{P_1V_2} \right)  \,, \quad
\mathrm{arg}\left( \widetilde{R}^{P_2V_1}\right) \,, \quad
\mathrm{arg}\left(\frac{ T^{P_2V_1}}{ T^{P_1V_2}} \right)\,,\label{eq:parameters1}
\end{align}
which do not depend on kinematic variables. 

As an example, we demonstrate the extraction of the above theory parameters in a minimal simplified scenario that consists of three approximations: 
\begin{enumerate}
\item The three body decay factorizes into two pseudo two-body decays (in the narrow width approximation), 
	\begin{align}	
	D^0	       &\rightarrow (V^{\pm}\rightarrow P^\pm P^0) P^{\mp}\,, &
 	\overline{D}^0 &\rightarrow (V^{\pm}\rightarrow P^\pm P^0) P^{\mp}\,,
	\end{align}
	where the decays $D^0\rightarrow V^{\pm} P^{\mp}$ and $\overline{D}^0\rightarrow V^{\pm} P^{\mp}$ are weak decays, and the decays $V^{\pm}\rightarrow P^\pm P^0$ are purely strong decays.
\item In principle, there is no need at this point to specify the amplitude for the propagator between the two decays. However, for illustration, we employ here a Breit-Wigner distribution, and do not take into account Blatt-Weisskopf factors that would account for the finite size of the $D^0$ and $\rho$. 
Our results are independent of this choice, and whenever a Breit-Wigner distribution is used, it can also be replaced by a more sophisticated distribution. 
\item For simplicity, we consider the interference of only two resonances, namely, $D^0\rightarrow \pi^+ ( \rho^- \rightarrow \pi^- \pi^0 )$ and $D^0\rightarrow \pi^- (\rho^+ \rightarrow \pi^+ \pi^0 )$. In the Dalitz plot region in which these two amplitudes interfere most strongly, the contributions from other resonances are small~\cite{Aubert:2008yd,Aaij:2014afa}. A more general treatment may include additional amplitudes and still allows extraction of all the physical parameters. 
\end{enumerate}
The above assumptions are useful approximations for the first round of measurements of the $\Delta U=0$ over $\Delta U=1$ matrix elements. They apply to measurements with large experimental relative uncertainties, in the range of 20\%--30\%. For future precision measurements, a more sophisticated and more complete Dalitz plot study is needed. 

These simplifications lead to the following contributions to the decay at a point 
\begin{align}
(s, t) \equiv (m^2_{\pi^-\pi^0},\, m^2_{\pi^+\pi^0}) 
\end{align}
in the Dalitz plot (where $m_X$ indicates the invariant mass of system $X$)~\cite{Kopp:2000gv, Aubert:2007ii}:
\begin{align}
\mathcal{A}\left( D^0\rightarrow \pi^- (\rho^+ \rightarrow \pi^0 \pi^+ )\right) &= 
		\mathcal{A}(D^0\rightarrow \pi^- \rho^+ )  
		\mathcal{BW}_{\rho}( s, t  )  \,
		\mathcal{A}(\rho^+ \rightarrow \pi^+ \pi^0) \,, \label{eq:narrowWidth-1}\\
\mathcal{A}\left( D^0\rightarrow  ( \rho^- \rightarrow \pi^- \pi^0 ) \pi^+\right) &= 
		\mathcal{A}(D^0\rightarrow  \rho^- \pi^+  ) 
		\mathcal{BW}_{\rho}( t, s )  \, 
		\mathcal{A}(\rho^- \rightarrow \pi^- \pi^0) \,, \label{eq:narrowWidth-2}
\end{align}
where the propagator
\begin{align}
\mathcal{BW}_{\rho}( s, t ) &\equiv	\frac{
			\left( m_{\pi^-\pi^+}^2 - s + \frac{(m_{D^0}^2 - m_{\pi^-}^2)(m_{\pi^0}^2 - m_{\pi^+}^2) }{m_{\rho^+}^2 }    \right)
		}{ 
			m_{\rho^+}^2 - t - i \Gamma_{\rho^+} m_{\rho^+}  
			}\,, \label{eq:breit-wigner}
\end{align}
includes a Breit-Wigner factor, and makes use of~\cite{Kopp:2000gv} 
\begin{align}
m_{\pi^-\pi^+}^2 &= m_{D^0}^2 + m_{\pi^-}^2 + m_{\pi^+}^2 + m_{\pi^0}^2 - s -  t \,. 
\end{align}
We emphasize that the amplitudes in Eqs.~(\ref{eq:narrowWidth-1}) and (\ref{eq:narrowWidth-2}) are complex numbers, and all dependences on the kinematical variables of the Dalitz plot are captured by the Breit-Wigner functions because of the model assumptions that we made.
CP invariance of QCD implies for the strong decays (in the phase convention we choose) 
\begin{align}
\mathcal{A}(\rho^+ \rightarrow \pi^+ \pi^0) &= \mathcal{A}(\rho^- \rightarrow \pi^- \pi^0)\,. \label{eq:rhopipi}
\end{align}
The complete Dalitz-plot amplitude reads then: 
\begin{align}
& \frac{ \mathcal{A}(D^0\rightarrow \pi^+\pi^-\pi^0 )}{\lambda_{sd}} =\frac{1}{\lambda_{sd}} \left( 
	\mathcal{A}\left( D^0\rightarrow \pi^- (\rho^+ \rightarrow \pi^0 \pi^+ )\right) +
	\mathcal{A}\left( D^0\rightarrow  ( \rho^- \rightarrow \pi^- \pi^0 ) \pi^+\right) 
		   \right) \nn\\ 
 &=   A_1( s, t ) \left(- 1  - \widetilde{\lambda}_b \, \widetilde{R}^{P_2V_1} \right) + 
	A_2( s , t )  
			\left(- 1 - \widetilde{\lambda}_b \, \widetilde{R}^{P_1V_2}\right) 
	\,,  \label{eq:Rho-exact}
\end{align}
where the hadronic functions
\begin{align}
A_1( s, t ) &\equiv \mathcal{BW}( s, t )		
				\mathcal{A}(\rho^+ \rightarrow \pi^+ \pi^0) T^{P_2V_1}\,, \\
A_2( s, t ) &\equiv \mathcal{BW}( t, s )  
			\mathcal{A}(\rho^- \rightarrow \pi^- \pi^0) T^{P_1V_2}
\end{align}
carry strong phases only. 

The parameters $T^{P_iV_j}$ and their relative phase can be extracted from the $\Gamma(D\rightarrow \pi^+\pi^-\pi^0)$ Dalitz plot, 
in which the contribution of 
$\widetilde{\lambda}_b$-terms is negligible:
\begin{align}
&\frac{\vert\mathcal{A}(D^0\rightarrow \pi^+\pi^-\pi^0 ) \vert^2}{\vert\lambda_{sd}\vert^2} =
\vert A_1( s , t )\vert^2 +
\vert A_2( s , t )\vert^2 +\nn\\
&\quad 2 \vert A_1( s , t )\vert \vert A_2( s, t )\vert \cos(\delta_{A_1}( s , t ) - \delta_{A_2}( s , t  ) )  +\mathcal{O}\left(\widetilde{\lambda}_b\right)\,, \label{eq:rate} 
\end{align}
where we use the shorthand notation for phases, 
\begin{align}
\delta_X &\equiv \mathrm{arg}(X)\,.
\end{align}
Writing the CP-conjugate amplitude 
\begin{align}
& \mathcal{A}(\overline{D}^0\rightarrow \pi^+\pi^-\pi^0 )/\lambda_{sd}^* = 
  	 A_1( s , t ) \left(- 1 - \widetilde{\lambda}_b^* \, \widetilde{R}^{P_2V_1} \right) +
 	  A_2( s , t ) \left(- 1 - \widetilde{\lambda}_b^* \, \widetilde{R}^{P_1V_2}\right)\,,
\end{align}
we can use the CP difference  
\begin{align}
& \frac{ \vert\mathcal{A}\vert^2 - \vert\overline{\mathcal{A}}\vert^2 }{  -4 \vert \lambda_{sd}\vert^2 \mathrm{Im}(\widetilde{\lambda_b}) }  =
\vert A_2( s , t  )\vert^2 \vert \widetilde{R}^{P_1V_2}\vert \sin( \delta_{ \widetilde{R}^{P_1V_2}  } ) - \nn\\ 
&\quad \vert A_1( s , t )\vert \vert A_2( s , t )\vert \vert \widetilde{R}^{P_1V_2}\vert 
\sin\left(\delta_{A_1}( s , t ) - \delta_{A_2}( s , t  ) - \delta_{\widetilde{R}^{P_1V_2}}\right) + \nn\\
&\quad \vert A_1( s , t  )\vert \vert A_2( s , t )\vert \vert \widetilde{R}^{P_2V_1}\vert 
\sin\left(\delta_{A_1}( s , t ) -  \delta_{A_2}( s , t ) + \delta_{ \widetilde{R}^{P_2V_1}  }\right) + \nn\\
&\quad \vert A_1( s , t  )\vert^2 \vert \widetilde{R}^{P_2V_1}\vert \sin(\delta_{ \widetilde{R}^{P_2V_1}  }) \label{eq:CPasym}
\end{align}
to extract $R^{P_iV_j}$ and their relative phase. 
Eq.~(\ref{eq:rhopipi}) implies that the phase of the strong decays $\rho\rightarrow \pi\pi$ cancels in the difference $\delta_{A_1} - \delta_{A_2}$. 
We also define the local direct CP asymmetry as
\begin{align}
a_{CP}^{\mathrm{dir}}(s,t) &\equiv \frac{
	\vert \mathcal{A}\vert^2 - \vert \overline{\mathcal{A}}\vert^2
	}{
	\vert \mathcal{A}\vert^2 + \vert \overline{\mathcal{A}}\vert^2
	}\,, 
	\label{eq:asymST}
\end{align} 
where we make the dependence on the Dalitz plot variables explicit on the LHS.
Fitting Eqs.~(\ref{eq:rate}), (\ref{eq:CPasym}) to the Dalitz-plot data enables the determination of all the parameters in Eq.~(\ref{eq:parameters1}). Consequently, we can determine the ratios $\widetilde{R}^{P_iV_j}$, \emph{i.e.}, investigate the $\Delta U=0$ rule for $D\rightarrow \rho\pi$. 

We emphasize the importance of the conceptual difference between two-body decays, such as $D\rightarrow \pi^+\pi^-$, and three-body decays such as $D\rightarrow \pi^+\pi^-\pi^0$, to which the pseudo two-body decays $D\rightarrow \rho \pi$ contribute. 
The three-body decays provide two advantages: first, interference between  
$D^0\rightarrow \pi^+ ( \rho^- \rightarrow \pi^- \pi^0 )$ 
and $D^0\rightarrow \pi^- (\rho^+ \rightarrow \pi^+ \pi^0 )$.
Second, kinematic dependence that enables studying the interference to obtain the magnitudes and phases of $\widetilde{R}^{P_1V_2}$ and $\widetilde{R}^{P_2V_1}$ from a time-integrated CP-asymmetry measurement. This is not possible for the corresponding parameters of $D\rightarrow \pi^+\pi^-$ and $D\rightarrow K^+K^-$, where time-dependent or quantum-correlated-production measurements are needed in order to completely solve the system.

\subsubsection{Cancellation of Production and Detection Asymmetry \label{sec:productionAsym}}

Measurement of a decay-rate asymmetry in 2-body $D$ decays is hampered by an experimental asymmetry, which arises from two main sources~\cite{Aaij:2019kcg}.
The first is due to a difference in the production rates for $D^0$ and $\overline{D}^0$ are different in $pp$ collisions. The second is that at LHCb, the detection efficiency of the soft pion from the $D^*$ decay used to tag the flavor of the $D^0$ and $\overline{D}^0$ is charge dependent. 
In order to study the effect of these production and detection asymmetries, we combine them in a parameter $\delta$ and write the observed local CP asymmetry as
\begin{align}
a_{CP}^{\mathrm{dir},\delta} &\equiv \frac{
	\vert A\vert^2 - (1-\delta) \vert \overline{A}\vert^2
	}{
	\vert A\vert^2 + (1-\delta) \vert \overline{A}\vert^2
	}\,. 
\end{align}
Expanding in $\delta$ and $\widetilde{\lambda}_b$, one obtains, for two-body decays,
\begin{align}
a_{CP}^{\mathrm{dir},\delta}(D^0\rightarrow P^+P^-) &=  a_{CP}^{\mathrm{dir}}(D^0\rightarrow P^+P^- ) + \frac{\delta}{2} 
	\left(1 + \mathcal{O}\left(\widetilde{\lambda}_b^2\right)\right)\,, 
\label{eq:obs-asym-PP}
\end{align}
i.e.~the production and detection asymmetry give a constant contribution to the observed asymmetry.
Since $\delta$ is independent of the decay mode of the $D$ meson, LHCb take the difference between $a_{CP}^{\mathrm{dir},\delta}(D^0\rightarrow \pi^+\pi^-)$ and $a_{CP}^{\mathrm{dir},\delta}(D^0\rightarrow K^+K^-)$ to extract a physical asymmetry.

This reasoning is valid also for the observed asymmetry in pseudo two-body decays within the Dalitz plot:
\begin{align}
a_{CP}^{\mathrm{dir},\delta}(D^0\rightarrow P^{\pm}V^{\mp})(s,t) &= a_{CP}^{\mathrm{dir}}(D^0\rightarrow P^{\pm}V^{\mp} )(s,t) + \frac{\delta}{2} 
	\left(1 + \mathcal{O}\left(\widetilde{\lambda}_b^2\right)(s,t)\right)\,.
\end{align}
However, in this case the physical parameters are obtained from a fit of the distribution of events in the Dalitz plot. Since the leading $\delta$ contribution is $(s,t)$-independent, it can be obtained from the integrated asymmetry, 
\begin{align}
\int a_{CP}^{\mathrm{dir},\delta}(D^0\rightarrow P^{\pm}V^{\mp})(s,t) ds\, dt\,, 
\end{align}
independently of the Dalitz-plot analysis. Therefore, $\delta$ does not directly impact the determination of the physical parameters, Eq.~(\ref{eq:parameters1}), and there is no need to subtract the asymmetries of different decay modes.

We have argued that, as in the case of two-body decays, the
production and detection asymmetries cancel, basically since they are
independent of the $D$ decay mode. There is, however, the possibility of an $(s,t)$-dependent  
detection asymmetry that is not present in the two-body
decay. 
This asymmetry arises from the fact that
in decays such as $D^0\rightarrow (K^{*+}\rightarrow K^+\pi^0) K^-$ and
$D^0\rightarrow (K^{*-}\rightarrow K^-\pi^0) K^+ $, the momenta of
the $K^+$ and $K^-$ are not the same.
We expect this detection asymmetry dependence on $(s,t)$ to be
small, since the detection efficiency depends primarily on the momentum in the laboratory frame.
Clearly, this systematic effect deserves further experimental study.
Yet, this is beyond the scope of this paper, where we focus on the
theoretical foundations of this measurement.

\subsubsection{Numerical Example \label{sec:scenario} }

\begin{figure}[b]
 \begin{center}
\subfigure[\quad $\widetilde{R}^{P_1V_2} =  \exp(i 3\pi/2 )$, \quad
		 $\widetilde{R}^{P_2V_1} =  \exp(i \pi/3 )$ ]{ \includegraphics[width=0.48\textwidth]{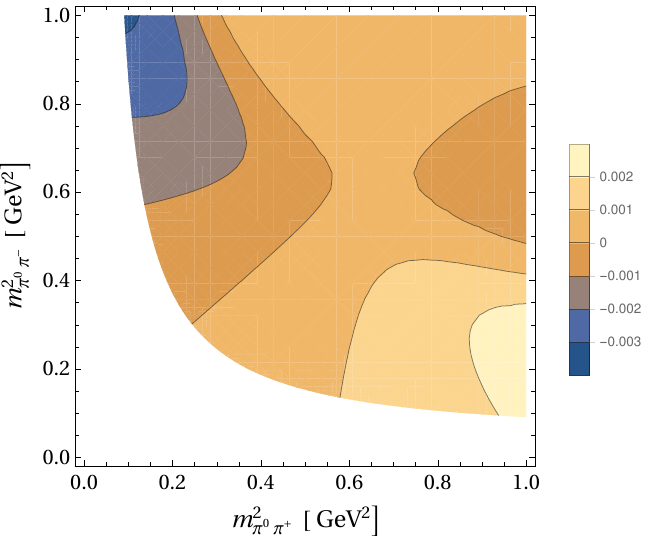} }  
\subfigure[\quad $\widetilde{R}^{P_1V_2} =  \exp(i \pi/2 )$, \quad
		 $\widetilde{R}^{P_2V_1} =  \exp(i \pi/3 )$ ]{ \includegraphics[width=0.48\textwidth]{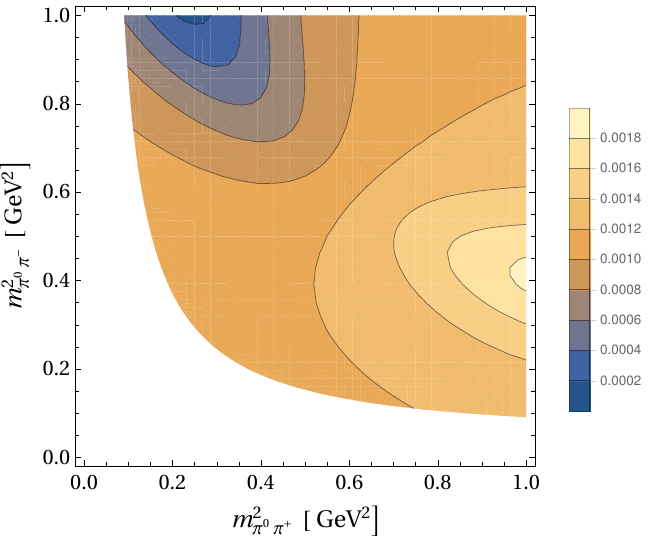} }  
\subfigure[\quad $\widetilde{R}^{P_1V_2} = \frac{1}{2} \exp(i \pi/2 )$,\quad 
		 $\widetilde{R}^{P_2V_1} =  \exp(i \pi/3 )$ ]{ \includegraphics[width=0.48\textwidth]{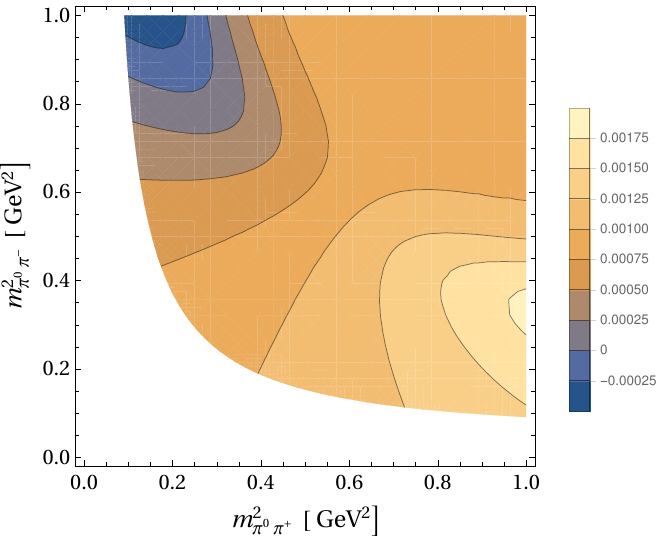} }  
\subfigure[\quad $\widetilde{R}^{P_1V_2} =  \exp(i \pi/2 )$,\quad  
		 $\widetilde{R}^{P_2V_1} = \frac{1}{4} \exp(i \pi/3 )$ ]{ \includegraphics[width=0.48\textwidth]{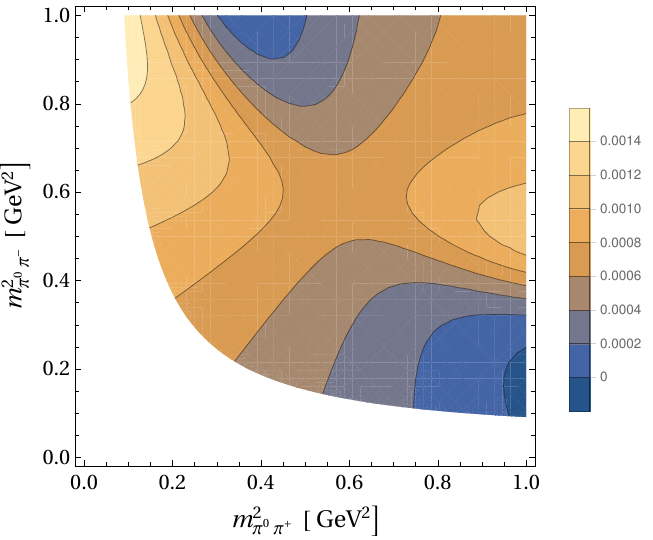} }  
  \caption{Local CP asymmetry of $D\rightarrow \pi^+\pi^-\pi^0$ in the region of the overlap of the $\rho^{\pm}$ resonances, for different values of $\widetilde{R}^{P_1V_2}$ and $\widetilde{R}^{P_2V_1}$.  \label{fig:plots}}
\end{center}
\end{figure}

For illustration, in Fig.~\ref{fig:plots} we show the local CP asymmetry of 
$D^0\rightarrow \pi^+\pi^-\pi^0$ for example values of $\widetilde{R}^{P_iV_j}$, using the numerical input from Ref.~\cite{PDG2020}.
Averaging results from hadron production and from $\tau$~decays/$e^+e^-$ production given therein, we take 
\begin{align}
m_{\rho^{\pm}} &= (774.36\pm 0.32) \, \mathrm{MeV}\,,  \\
\Gamma_{\rho^{\pm}} &= (149.21\pm 0.76)\, \mathrm{MeV}\,. 
\end{align}

We neglect $\mathcal{O}(\widetilde{\lambda}_b)$ in the estimation of $\vert T^{P_iV_j}\vert$, using
\begin{align}
\vert T^{P_1 V_2}\vert &= \sqrt{\frac{1}{\lambda_{sd}^2}\frac{\mathcal{B}(D^0\rightarrow \rho^-\pi^+)}{\mathcal{P}(D^0, \rho^-,\pi^+)}}\,, \label{eq:extraction-1}\\
\vert T^{P_2 V_1}\vert &= \sqrt{\frac{1}{\lambda_{sd}^2}\frac{\mathcal{B}(D^0\rightarrow \rho^+\pi^-)}{\mathcal{P}(D^0, \rho^+,\pi^-)}}\,, \label{eq:extraction-2}
\end{align}
with the phase space factor
\begin{align}
\mathcal{P}(D, \rho, \pi) &= \tau_D \times \frac{1}{16\pi m_D^3} 
	\sqrt{   
	\left(m_D^2 - ( m_{\rho} - m_{\pi} )^2 \right)
	\left(m_D^2 - ( m_{\rho} + m_{\pi} )^2 \right)
	}\,.
\end{align}
Furthermore, for our numerical example we choose $T^{P_1 V_2}$ and $T^{P_2 V_1}$ to be real. This is somewhat motivated from the leading contribution of a $1/N_c$--expansion of the tree diagram~\cite{Buras:1985xv, Muller:2015lua}. However, additional diagrams may be sizable, invalidating this assumption.

The results in Fig.~\ref{fig:plots} illustrate how the Dalitz plot CP asymmetry depends on the rescattering amplitudes. Locally, effects at the per mill level are to be expected for order-one rescattering, in agreement with what is seen in $D\rightarrow PP$ decays. 

The four examples of numerical values of $\widetilde{R}^{P_1V_2}$ and $\widetilde{R}^{P_2V_1}$ correspond to different scenarios of $\mathcal{O}(1)$ rescattering. This value is motivated by the measurement of $\Delta a_{CP}^{\mathrm{dir}}$ in 
two-body decays~\cite{Grossman:2019xcj}.
The figures illustrate that the shape of the Dalitz plot is indeed sensitive to the specific ratio of rescattering amplitudes compared to tree amplitudes, enabling a future extraction of these ratios by experiment.


\section{Relation to $D^0\rightarrow K^+K^-\pi^0$ \label{sec:breaking}}

\subsection{Amplitude Parametrization}

The decays 
\begin{align}\label{eq:Dpipipi}
D^0\rightarrow \pi^\pm \rho^\mp  \rightarrow \pi^+\pi^-\pi^0\, .
\end{align} 
are related by $U$-spin to
\begin{align}\label{eq:DKKpi}
D^0\rightarrow K^\pm K^{*\mp} \rightarrow K^+ K^-\pi^0.
\end{align}
The LHCb discovery of CP violation in two-body decays~\cite{Aaij:2019kcg} employed an analogous $U$-spin relation, between $D^0\rightarrow \pi^+ \pi^-$ and $D^0\rightarrow K^+ K^-$.
The motivation to consider CP violation in both modes together, and specifically their difference, is twofold. First, experimentally, this leads to cancellation of the production and detection asymmetry and of systematic uncertainties in the two-body modes. For the case of three-body decays, as we discuss in Sec.~\ref{sec:productionAsym}, the production and detection asymmetry is $(s,t)$-independent and hence does not impact the extraction of the physical parameters. 
The question of whether additional experimental systematic uncertainties cancel in the Dalitz-plot asymmetry is beyond the scope of this paper and we do not discuss it further.  
A second advantage in the case of two-body decays is that the two modes are related by a complete interchange of all $d$ and $s$ quarks, resulting in~\cite{Gronau:2000zy} 
\begin{align}
	a_{CP}^{\mathrm{dir}}(D^0\to\pi^+\pi^-) = -a_{CP}^{\mathrm{dir}}(D^0\to K^+K^-).
	\label{eq:equal-opposite-ACPs}
\end{align}
Therefore, the difference of these asymmetries is twice as large as each one separately, and is thus measured with a higher statistical significance.

All this motivates considering analogous differences in three-body decays as well.
In order to explore the relation between the modes of Eqs.~(\ref{eq:Dpipipi}) and~(\ref{eq:DKKpi}), we would like to consider the full set of $D^0\to V^\pm P^\mp$ amplitudes within a $U$-spin expansion in full generality,~\emph{i.e.}~incorporating also possible New Physics (NP) amplitudes. In doing so, we note that 
full $U$-spin multiplets include also Cabibbo-favored (CF) and doubly Cabibbo-suppressed (DCS) amplitudes. Within the SM, these have different structures from that of the singly Cabibbo-suppressed (SCS) amplitudes. 
It is therefore important to consider carefully how to generalize the corresponding SM parametrizations to include NP effects. 

The decay amplitudes for SCS decays have a structure of two terms with a non-zero relative weak phase. 
As mentioned above, any third term with an additional relative weak and strong phase can be absorbed by redefinition into the other parameters~\cite{Grossman:1997gd}. Therefore, the parametrizations employed above are sufficient also when considering any NP amplitudes.
However, for CF and DCS decays, a second term with a relative weak phase arises in the SM only at the second order in the weak interaction, \emph{i.e.}~with two $W$-boson exchanges, which we neglect. Consequently, in order to incorporate potential NP amplitudes in CF and DCS decays, we need to introduce additional terms with a relative weak phase. Note, however, that the $U$-spin transformation behavior of such an additional term is the same as that of the leading (SM) term.
 
The full set of $D^0\to V^\pm P^\mp$ amplitudes is then altogether given by
\begin{align}
\mathcal{A}( D^0\rightarrow \pi^+ K^{*-} ) &= V_{cs}^* V_{ud} \left( t_0^{P_1V_2} - \frac{1}{2} t^{P_1V_2}_1 \right) + \lambda_{\mathrm{NP}}^{\mathrm{CF}} r_{\mathrm{NP}, \mathrm{CF}}^{P_1V_2}\,, \label{eq:uspin-1}\\
\mathcal{A}( D^0\rightarrow \pi^+ \rho^- )   &= - \lambda_{sd}\, \left( t_0^{P_1V_2} + s_1^{P_1V_2} + \frac{1}{2} t_2^{P_1V_2} \right) - \lambda_b\, \left(r_0^{P_1V_2} - \frac{1}{2} r_1^{P_1V_2}\right)\,, \label{eq:uspin-2}\\
\mathcal{A}( D^0\rightarrow K^+ K^{*-}   )    &=   \lambda_{sd}\, \left( t_0^{P_1V_2} - s_1^{P_1V_2} + \frac{1}{2} t_2^{P_1V_2} \right) -  \lambda_b\, \left(r_0^{P_1V_2} + \frac{1}{2} r_1^{P_1V_2} \right)\,, \label{eq:uspin-3} \\ 
\mathcal{A}( D^0\rightarrow K^+ \rho^-   )	  &= V_{cd}^* V_{us} \left(t_0^{P_1V_2} + \frac{1}{2} t_1^{P_1V_2} \right) + \lambda_{\mathrm{NP}}^{\mathrm{DCS}} r_{\mathrm{NP}, \mathrm{CF}}^{P_1V_2}\,, \label{eq:uspin-4}
\end{align}
and
\begin{align}
\mathcal{A}( D^0\rightarrow  K^- \rho^+   )    &= V_{cs}^* V_{ud} \left( t_0^{P_2V_1} - \frac{1}{2} t^{P_2V_1}_1 \right) + \lambda_{\mathrm{NP}}^{\mathrm{CF}} r_{\mathrm{NP}, \mathrm{CF}}^{P_2V_1}\,, \label{eq:uspin-5} \\ 
\mathcal{A}( D^0\rightarrow \pi^- \rho^+ )   & = - \lambda_{sd}\, \left( t_0^{P_2V_1} + s_1^{P_2V_1} + \frac{1}{2} t_2^{P_2V_1} \right) - \lambda_b\, \left(r_0^{P_2V_1} - \frac{1}{2} r_1^{P_2V_1}\right) \,, \label{eq:uspin-6} \\
\mathcal{A}( D^0\rightarrow K^- K^{*+}   )  &=   \lambda_{sd}\, \left( t_0^{P_2V_1} - s_1^{P_2V_1} + \frac{1}{2} t_2^{P_2V_1} \right) - \lambda_b\, \left(r_0^{P_2V_1} + \frac{1}{2} r_1^{P_2V_1} \right)\,, \label{eq:uspin-7} \\
\mathcal{A}( D^0\rightarrow \pi^- K^{*+}  )  &=  V_{cd}^* V_{us} \left(t_0^{P_2V_1} + \frac{1}{2} t_1^{P_2V_1} \right) + \lambda_{\mathrm{NP}}^{\mathrm{DCS}} r_{\mathrm{NP}, \mathrm{CF}}^{P_2V_1} \,. \label{eq:uspin-8} 
\end{align}  
Note that the above parametrization serves two purposes. First, it can be used as a completely model-independent parametrization. 
Second, the parameters can be interpreted as part of a $U$-spin expansion with a power counting, such that the subscripts \lq\lq{}0\rq\rq{}, \lq\lq{}1\rq\rq{} and \lq\lq{}2\rq\rq{} indicate the $U$-spin limit, first-order $U$-spin  breaking, and second-order $U$-spin breaking, respectively.
In this section we use this $U$-spin power-counting interpretation. 

Note further that this parametrization is in complete analogy to the following generalization of the notation of   
Refs.~\cite{Brod:2012ud,Grossman:2019xcj} for the two-body decays $D^0\rightarrow P^+P^-$: 
\begin{align}
\mathcal{A}(D^0\rightarrow K^- \pi^+ )     &= V_{cs}^* V_{ud} \left( t_0 - \frac{1}{2} t_1 \right) + 
					\lambda_{\mathrm{NP}}^{\mathrm{CF}} \, r_{\mathrm{NP}}^{\mathrm{CF}}\,, \label{eq:Dpipi-1}\\
\mathcal{A}( D^0\rightarrow \pi^+ \pi^- )  &= - \lambda_{sd}\, \left( t_0 + s_1 + \frac{1}{2} t_2 \right) - 
						 \lambda_b\, \left(r_0 - \frac{1}{2} r_1\right)\,, \label{eq:Dpipi-r}\\
\mathcal{A}( D^0\rightarrow K^+ K^- )      &=   \lambda_{sd}\, \left( t_0 - s_1 + \frac{1}{2} t_2 \right) - 
						 \lambda_b\, \left(r_0 + \frac{1}{2} r_1 \right)\,, \\ 
\mathcal{A}(D^0\rightarrow K^+ \pi^- )	   &= V_{cd}^* V_{us} \left(t_0 + \frac{1}{2} t_1 \right) +
					 \lambda_{\mathrm{NP}}^{\mathrm{DCS}} \, r_{\mathrm{NP}}^{\mathrm{DCS}}\,. \label{eq:Dpipi-4} 
\end{align}
In Appx.~\ref{sec:parametrization} we list the complete set of observables that correspond to the amplitudes of Eqs.~(\ref{eq:uspin-1})--(\ref{eq:Dpipi-4}).

We use the following notation for the ratio of matrix elements, similarly to Eq.~(\ref{eq:r-ratio}),
\begin{align}
	\widetilde{r_0}^{P_iV_j} \equiv \frac{r_0^{P_iV_j}}{t_0^{P_iV_j}}, \quad \widetilde{r_1}^{P_iV_j} \equiv \frac{r_1^{P_iV_j}}{t_0^{P_iV_j}},\quad \widetilde{s_1}^{P_iV_j} \equiv \frac{s_1^{P_iV_j}}{t_0^{P_iV_j}}.
\end{align}
To first order in $U$-spin breaking, the amplitudes of the three-body decays are then given by 
\begin{align}
& \frac{ \mathcal{A}(D^0\rightarrow \pi^+\pi^-\pi^0 )}{\lambda_{sd}} \equiv  \frac{\mathcal{A}_{\pi\pi\pi}}{\lambda_{sd} } \nn\\ &=\frac{1}{\lambda_{sd}} \left( 
	\mathcal{A}\left( D^0\rightarrow \pi^- (\rho^+ \rightarrow \pi^0 \pi^+ )\right) +
	\mathcal{A}\left( D^0\rightarrow  ( \rho^- \rightarrow \pi^- \pi^0 ) \pi^+\right) 
		   \right) \\ \nn
 &=   \mathcal{BW}_{\rho}( m_{\pi^-\pi^0}^2,  m_{\pi^+\pi^0}^2) \mathcal{A}(\rho^{+} \rightarrow \pi^+ \pi^0)  t_0^{P_2V_1} \left(- 1 - \widetilde{s}_1^{P_2V_1} - \widetilde{\lambda}_b \,( \widetilde{r}_0^{P_2V_1}-\frac{1}{2}\widetilde{r}_1^{P_2V_1}) \right) \\ 
	&\quad + \mathcal{BW}_{\rho}( m_{\pi^+\pi^0}^2,  m_{\pi^-\pi^0}^2) \mathcal{A}(\rho^{-} \rightarrow \pi^- \pi^0)  t_0^{P_2V_1}  
			\left(- 1 - \widetilde{s}_1^{P_1V_2}- \widetilde{\lambda}_b \,( \widetilde{r}_0^{P_1V_2}-\frac{1}{2}\widetilde{r}_1^{P_1V_2})\right) 
	\,, \label{eq:Uspin1}\\\nn\\ 
& \frac{\mathcal{A}(D^0\rightarrow K^+K^-\pi^0 )}{\lambda_{sd}} \equiv  \frac{\mathcal{A}_{KK\pi}}{\lambda_{sd} } \nn\\
&=\frac{1}{\lambda_{sd}} \left( 
	\mathcal{A}\left( D^0\rightarrow K^- (K^{*+} \rightarrow \pi^0 K^+ )\right) +
	\mathcal{A}\left( D^0\rightarrow  ( K^{*-} \rightarrow K^- \pi^0 ) K^+\right) 
		   \right) \\ 
& = 	\mathcal{BW}_{K^*}( m_{K^-\pi^0}^2,  m_{K^+\pi^0}^2) \mathcal{A}(K^{*+} \rightarrow K^+ \pi^0)  t_0^{P_2V_1} \left(1 - \widetilde{s}_1^{P_2V_1} - \widetilde{\lambda}_b \, (\widetilde{r}_0^{P_2V_1}+\frac{1}{2}\widetilde{r}_1^{P_2V_1}) \right)  \nn\\ 
	&\quad + \mathcal{BW}_{K^*}(  m_{K^+\pi^0}^2, m_{K^-\pi^0}^2 )  
			\mathcal{A}(K^{*-} \rightarrow K^- \pi^0) t_0^{P_1V_2} \left(1- \widetilde{s}_1^{P_1V_2} - \widetilde{\lambda}_b \, (\widetilde{r}_0^{P_1V_2}+\frac{1}{2}\widetilde{r}_1^{P_1V_2})\right)\,,  \label{eq:Uspin2}
\end{align}
with $\mathcal{BW}_{K^*}( m_{K^-\pi^0}^2,  m_{K^+\pi^0}^2 )$ defined analogously to Eq.~(\ref{eq:breit-wigner}). 
As in the case of the $\rho^\pm$ strong decays, CP invariance of QCD implies 
\begin{align}
\mathcal{A}(K^{*+} \rightarrow K^+ \pi^0) &= \mathcal{A}(K^{*-} \rightarrow K^- \pi^0)\,
\end{align}
(up to an arbitrary choice of phase).

In order to discuss the kinematic $U$-spin breaking between 
$\mathcal{A}(D^0\rightarrow K^+K^-\pi^0)$ and $\mathcal{A}(D^0\rightarrow \pi^+\pi^-\pi^0) $ related to the Breit-Wigner functions, we define 
\begin{align}
        M_V &\equiv \frac{m_{\rho^{\pm}} + m_{K^{*\pm}}}{2}, &  
	\Delta_{M_V} &\equiv \frac{m_{K^{*\pm}} - m_{\rho^\pm}}{2M_V}\,,\\ 
        M_P &\equiv \frac{m_{\pi^{\pm}} + m_{K^{\pm}}}{2}, &  
	\Delta_{M_P} &\equiv \frac{m_{K^{\pm}} - m_{\pi^\pm}}{2M_P}\,,\\ 
        \Gamma &\equiv  \frac{\Gamma_{\rho^\pm} +\Gamma_{K^{*\pm}}}{2}, & 
	\Delta_\Gamma &\equiv \frac{\Gamma_{K^{*\pm}}-\Gamma_{\rho^{\pm}}}{2\Gamma},
\end{align}
so that 
\begin{align}
m_{\rho^{\pm}}      &= M_V (1 - \Delta_{M_V} )\,, &
m_{K^{*\pm}}        &= M_V (1 + \Delta_{M_V} ) \,, \\
m_{\pi^{\pm}}      &= M_P (1 - \Delta_{M_P} )\,, &
m_{K^{\pm}}        &= M_P (1 + \Delta_{M_P} ) \,, \\
\Gamma_{\rho^{\pm}} &= \Gamma (1 - \Delta_{\Gamma}) \,, &
\Gamma_{K^{*\pm}}   &= \Gamma (1 + \Delta_{\Gamma}) \,. 
\end{align}
Numerically, we have 
\begin{align}
\Delta_{M_V}  \approx  0.07\,, \qquad 
\Delta_{M_P}  \approx  0.56\,, \qquad
\Delta_\Gamma \approx -0.50\,, \label{eq:deltaGamma}
\end{align}
\emph{i.e.}~the kinematic $U$-spin breaking is large.
The Breit-Wigner functions then have the following form,
\begin{align}
\mathcal{BW}_{\rho}(s,t) &= \mathcal{BW}(s,t)+\Delta_\mathcal{BW}^\rho(s,t)\,, \\ \nonumber
\mathcal{BW}_{K^*}(s,t) &= 	\mathcal{BW}(s,t)+\Delta_\mathcal{BW}^{K^*}(s,t)\,,
\end{align}
where $\Delta_\mathcal{BW}^{\rho,K^*}$ are the leading order corrections to the U-spin limit, and 
where we use $s$ and $t$ to denote the $U$-spin limit variables
\begin{align}
	s\equiv m_{12}^2, \quad t\equiv m_{23}^2\,,
\end{align}
and the subscript indices 1, 2, and 3 correspond to the positively charged, negatively charged, and neutral final state pseudoscalars, respectively. 

The $U$-spin limit propagator is given by
\begin{align}
\mathcal{BW}(s,t) &= \frac{-3 M_P^2-m_{D^0}^2+2 s+t}{ -M_V^2 + t + i \Gamma M_V}\equiv |\mathcal{BW}(s,t)|e^{i\delta_\mathcal{BW}(s,t)}\,.
\end{align}
We do not take into account here the $U$-spin breaking that comes from the fact that these phase space variables themselves
are different, \emph{i.e.}  
\begin{align}
m_{K^-\pi^0}^2 &\neq m_{\pi^-\pi^0}^2\,,  \qquad
m_{K^+\pi^0}^2 \neq m_{\pi^+\pi^0}^2\,,
\end{align}
nor from the fact that the kinematic boundaries of the Dalitz plots are different.

A fundamental problem is that there is no obvious way to meaningfully associate two points of different Dalitz plots of $U$-spin-related decays. 
Therefore, we do not know how large the corresponding $U$-spin breaking effects are.
Neglecting these effects is clearly a rough estimate.
Still, we expect that there exist regions of the Dalitz plot that are large enough that integrating over each such region has a $U$-spin-breaking effect similar to that of the total rate, namely at the nominal size of $m_s/\Lambda_{\mathrm{QCD}}\sim 30\%$ that is also found in $D\rightarrow PP$ decays~\cite{Hiller:2012xm, Muller:2015lua, Grossman:2019xcj}. The sizes and placements of such regions are unknown at this time, and we proceed with the discussion without addressing this problem in detail.

In analogy to
\begin{align}
\Delta a_{CP}^{P,\, \mathrm{dir}} &\equiv a_{CP}^{\mathrm{dir}}(D^0\rightarrow K^+K^-) - a_{CP}^{\mathrm{dir}}(D^0\rightarrow \pi^+\pi^-)\,, \\
\Sigma a_{CP}^{P,\, \mathrm{dir}} &\equiv a_{CP}^{\mathrm{dir}}(D^0\rightarrow K^+K^-) + a_{CP}^{\mathrm{dir}}(D^0\rightarrow \pi^+\pi^-)\,, 
\end{align}
we consider the corresponding difference and sum for three-body decays, 
\begin{align}
\Delta a_{CP}^{V,\, \mathrm{dir}}(s,t) &\equiv \Big(\frac{|{\cal A}_{KK\pi}(s,t)|^2-|\bar{\cal A}_{KK\pi}(s,t)|^2}{|{\cal A}_{{K^*}\to K\pi}|^2} - \frac{|{\cal A}_{\pi\pi\pi}(s,t)|^2-|\bar{\cal A}_{\pi\pi\pi}(s,t)|^2}{|{\cal A}_{{\rho}\to \pi\pi}|^2}\Big)\,, \\
\Sigma a_{CP}^{V,\, \mathrm{dir}}(s,t) &\equiv \Big(\frac{|{\cal A}_{KK\pi}(s,t)|^2-|\bar{\cal A}_{KK\pi}(s,t)|^2}{|{\cal A}_{{K^*}\to K\pi}|^2} + \frac{|{\cal A}_{\pi\pi\pi}(s,t)|^2-|\bar{\cal A}_{\pi\pi\pi}(s,t)|^2}{|{\cal A}_{{\rho}\to \pi\pi}|^2}\Big)\,,
\end{align}
which are normalized to the strong decay amplitude squared.

\subsection{Approximate CP Asymmetry Sum Rule}

The CP asymmetries of the three-body decay modes differ from the analogous two-body modes in several ways.
Naively, one would think that the addition of a $\pi^0$ to the final states
of $D^0\rightarrow \pi^+\pi^-$ and $D^0\rightarrow K^+K^-$ is not important.
Yet, the key difference is 
that the three-body decays $D^0\rightarrow \pi^+\pi^-\pi^0$ and
$D^0\rightarrow K^+K^-\pi^0$ are not connected by a complete
interchange of $d$ and $s$ quarks. Consequently, the theorem of
Ref.~\cite{Gronau:2000zy}, which connects CP asymmetries of decays
with a complete interchange of $d$ and $s$ quarks, does not apply~\cite{Grossman:2018ptn}. 
In particular, there is no exact 
$U$-spin relation between the CP asymmetries in $D^0\rightarrow \pi^+\pi^-\pi^0$ and in
$D^0\rightarrow K^+K^-\pi^0$.

The three-body analogy to the two-body system is also not perfect when considering deviations from the $U$-spin limit. While the $U$-spin breaking 
in the two-body case manifests entirely in the $U$-spin breaking parameters of the amplitude, the three-body decays have an additional source of breaking, which arises from the propagators of the intermediate resonances, \emph{i.e.}, the different masses and widths of the $K^*$ and $\rho$.

While, as we mention above, there is no $U$-spin relation between the CP asymmetries in the  $D^0\rightarrow \pi^+\pi^-\pi^0$ and 
$D^0\rightarrow K^+K^-\pi^0$ three-body modes, there is a relation between the underlying pseudo two-body decays
\begin{align}
& D^0\rightarrow \pi^+ \rho^-,\quad D^0\rightarrow K^+ K^{*-}\,,  \\ 
 \text{and} \quad  & D^0\rightarrow \pi^-\rho^+,\quad D^0\rightarrow K^- K^{*+}\,,
\end{align}
which are connected by a complete interchange of all $d$ and $s$
quarks, so that the theorem of Ref.~\cite{Gronau:2000zy} applies. 

We now make the approximations 
that the only relevant contributions to the three-body decays are from the 
$\rho^+(K^{*+})$ and $\rho^-(K^{*-})$ resonances, that the narrow-width approximation
holds for these resonances, and that we work to leading 
order in the $U$-spin expansion. Under these assumptions we have
the approximate sum rule
\begin{align}
\Sigma a_{CP}^{V,\, \mathrm{dir}}(s,t) = 0\,, \label{eq:sum-rule-V}
\end{align} 
corresponding to the analogous sum rule $\Sigma a_{CP}^{P,\,\mathrm{dir}} = 0$ in the two-body case.
The validity of Eq.~(\ref{eq:sum-rule-V}) can be used as a test of the
validity of our simplifying assumptions and the size of $U$-spin breaking. 

Considering $\Delta a_{CP}^{V,\, \mathrm{dir}}$, we find the $U$-spin limit,
\begin{align}
& \frac{\Delta a_{CP}^{V,\, \mathrm{dir}}(s,t)}{8 \vert \lambda_{sd}\vert^2 \mathrm{Im}\left(\widetilde{\lambda}_b \right)   } = \vert\mathcal{BW}(t,s)\vert^2 (t_0^{P_1V_2})^2 \vert \widetilde{r}_0^{P_1V_2}\vert \sin(\delta_{ \widetilde{r}_0^{P_1V_2} }) - \nn\\
& \vert\mathcal{BW}(s,t) \mathcal{BW}(t,s)\vert t_0^{P_2V_1} t_0^{P_1V_2}  \vert \widetilde{r}_0^{P_1V_2}\vert 
		\sin[\delta_{\mathcal{BW}(s,t)} - \delta_{\mathcal{BW}(t,s)} + \delta_{t_0^{P_2V_1}} - \delta_{t_0^{P_1V_2}}  - \delta_{ \widetilde{r}_0^{P_1V_2} }] + \nn \\ 
&\vert\mathcal{BW}(s,t) \mathcal{BW}(t,s)\vert  t_0^{P_2V_1}   t_0^{P_1V_2}  \vert \widetilde{r}_0^{P_2V_1} \vert 
	\sin(\delta_{\mathcal{BW}(s,t)} - \delta_{\mathcal{BW}(t,s)}  + \delta_{t_0^{P_2V_1}} - \delta_{t_0^{P_1V_2}} + \delta_{ \widetilde{r}_0^{P_2V_1} })+ \nn \\
&\vert\mathcal{BW}(s,t)\vert^2 (t_0^{P_2V_1})^2  \vert \widetilde{r}_0^{P_2V_1}\vert  \sin(\delta_{ \widetilde{r}_0^{P_2V_1} })\,. \label{eq:deltaACPV} 
\end{align}
We see that this equals twice the CP asymmetry in Eq.~(\ref{eq:CPasym}). This is a manifestation of the same feature that the two-body $\Delta A_{CP}^P$ measurements profit from, namely, that in the $U$-spin limit the asymmetries for the two modes have equal magnitudes and opposite signs, Eq.~(\ref{eq:equal-opposite-ACPs}).

\section{Conclusions \label{sec:conclusions}}

We show how future Dalitz-plot analyses of three-body charm decays can be used in order 
to extract the ratio of $\Delta U=0$ over $\Delta U=1$ matrix elements from the interference region of resonances. 
In particular, the Dalitz plot analyses allow for the extraction
of the relative strong phase with time-integrated CP violation measurements, without the need for time-dependent analysis or quantum correlations in $D$-pair production.
We present numerical examples of the local CP asymmetry for $D^0\rightarrow \pi^+\pi^-\pi^0$. For future reference, we also  
present the complete set of observables of $D^0\rightarrow V^{\pm} P^{\mp}$ decays.
We discuss the possibility of a combined Dalitz-plot analysis of  $D^0\rightarrow \pi^+\pi^-\pi^0$ and $D^0\rightarrow K^+K^-\pi^0$ in the region 
of charged resonances and the differences between $U$-spin related three-body modes relative to the case of pseudo two-body modes. 

To simplify the study of interference effects, we work in an approximate setup, in which the Dalitz-plot region of interest is dominated by the contributions of two narrow resonances. In the context of probing the existence of ${\cal O}(1)$ rescattering effects, such a rough framework suffices. Hopefully, with improved experimental precision in the future, our method will break down, and a more realistic approach will be needed. 
Such an approach will include a larger number of resonances and more precise descriptions of the resonant and non-resonant amplitudes. Such descriptions have already been used to describe the Dalitz-plot distributions for a number of $D$-meson decays.

\begin{acknowledgments}
The work of AD is partially supported by the Israeli council for higher education postdoctoral fellowship for women in science.
The work of YG is supported in part by the NSF grant PHY1316222.
S.S. is supported by a Stephen Hawking Fellowship from UKRI under reference EP/T01623X/1.
AS is supported by grants from ISF (2871/19, 2476/17), GIF (I-67-303.7-2015), BSF (2016113), MOST (Israel) (3-16543), and by the European Unions Horizon 2020 research and innovation programme under the Marie Skodowska-Curie grant agreement No. 822070.
\end{acknowledgments}

\begin{appendix}

\section{Parametrization of Two-Body Charm Decays \label{sec:parametrization}}

In this appendix, we present the independent observables of the 
complete $D^0\rightarrow P^+P^-$ and $D^0\rightarrow P^{\pm}V^{\mp}$ systems, including CF, SCS and DCS decays.

\subsection{Observables for $D^0\rightarrow P^{\pm}P^{\mp}$ Decays, $P=K,\pi$ \label{sec:appendixDPP}}

\begin{table}[t]
\begin{footnotesize}
\begin{center}
\begin{tabular}{l|c|c|c}
\hline
\multicolumn{2}{c|}{Observables}  & \multicolumn{1}{p{3cm}|}{Relations in the SM limit, $r_{\mathrm{NP}} = 0$} &\multicolumn{1}{p{3cm}}{Relations in the CP limit, $\mathrm{arg}(\lambda_{b}) = 0 $}   \\\hline\hline
\multirow{4}{3.3cm}{CP-averaged branching ratios} &
$\mathcal{B}( D\rightarrow K^- \pi^+  )$  &  &  \\
& $\mathcal{B}( D\rightarrow \pi^+ \pi^-)$ &   &  \\
& $\mathcal{B}( D\rightarrow K^+ K^-    )$ &   &  \\
& $\mathcal{B}( D\rightarrow K^+ \pi^-  )$ &   &  \\\hline
\multirow{4}{3.3cm}{Direct CP asymmetries} &
  $a_{CP}^{\mathrm{dir}}(D\rightarrow K^- \pi^+   )$ &  $=0$ & $=0$ \\
& $a_{CP}^{\mathrm{dir}}(D\rightarrow \pi^+ \pi^- )$ &  & $=0$ \\
& $a_{CP}^{\mathrm{dir}}(D\rightarrow K^+ K^-     )$ &  & $=0$  \\
& $a_{CP}^{\mathrm{dir}}(D\rightarrow K^+ \pi^-   )$ & $=0$  & $=0$ \\\hline
\multirow{4}{3.3cm}{Phases from interference effects} &
$\,\mathrm{arg}\left(\frac{\mathcal{A}(\overline{D}^0\rightarrow K^- \pi^+ )  }{\mathcal{A}(D^0\rightarrow K^- \pi^+   )} \right)$ 
	& $=\mathrm{arg}\left(\frac{  \mathcal{A}(D^0\rightarrow K^+ \pi^-) }{ \mathcal{A}(D^0\rightarrow K^- \pi^+ ) } \right)$  
	& $=\mathrm{arg}\left(\frac{  \mathcal{A}(D^0\rightarrow K^+ \pi^-) }{ \mathcal{A}(D^0\rightarrow K^- \pi^+ ) } \right)$ \\
& $\,\mathrm{arg}\left(\frac{\mathcal{A}(\overline{D}^0\rightarrow \pi^+ \pi^-) }{\mathcal{A}(D^0\rightarrow \pi^+ \pi^- )} \right)$ &  & $=0$  \\
& $\,\,\mathrm{arg}\left(\frac{\mathcal{A}(\overline{D}^0\rightarrow K^+ K^- )    }{\mathcal{A}(D^0\rightarrow K^+ K^- )    } \right)$ &  & $=0$  \\\hline
\# of independent observables \,\, &
\, 11 &  \, 9 & \,5   \\\hline\hline
\end{tabular}
\caption{Available observables in the $D^0\rightarrow P^+P^-$ system, and relations between them in the SM limit and in the CP limit.
\label{tab:obsDPP}}
\end{center}
\end{footnotesize}
\end{table}

The $D^0\rightarrow P^+P^-$ system, described with the general parametrization in Eqs.~(\ref{eq:Dpipi-1})--(\ref{eq:Dpipi-4}), has 12 independent observables, which we show in Table~\ref{tab:obsDPP}.
In addition to the CP limit, we consider also the SM limit
\begin{align}
\lambda_{\mathrm{NP}}^{\mathrm{CF}} &= \lambda_{\mathrm{NP}}^{\mathrm{DCS}} = 0\,. \label{eq:sm-limit}
\end{align} 
As discussed in Sec.~\ref{sec:breaking}, Eq.~(\ref{eq:sm-limit}) implies vanishing CP violation in the decays $D^0\rightarrow K^{\mp} \pi^{\pm}$
because in the SM, contributions to a relative weak phase in these decays come only at the second order 
in the weak interaction.
Another consequence of Eq.~(\ref{eq:sm-limit}) is the experimental sensitivity to the relative strong phase between the amplitudes of the decays 
$D^0\rightarrow K^{\mp} \pi^{\pm}$~\cite{Soffer:1998un}.
Consequently, altogether there are four strong phases, which in the SM are reduced to three, and in the CP limit to one.

Note that the phases
\begin{align}
& \mathrm{arg}\left(\frac{D^0\rightarrow K^+K^-}{D^0\rightarrow \pi^+\pi^-}\right)\,, \quad
 \mathrm{arg}\left(\frac{D^0\rightarrow K^-\pi^+}{D^0\rightarrow \pi^+\pi^-}\right)\,,
\end{align}
are inaccessible.
Note further that the phase
\begin{align}
\mathrm{arg}\left(\frac{  \mathcal{A}(D^0\rightarrow K^+ \pi^-) }{ \mathcal{A}(D^0\rightarrow K^- \pi^+ ) } \right) 
\end{align}
only becomes accessible when we assume that the CF and DCS decays do not violate CP, \emph{i.e.}~in the SM or CP limit. In that case we have 
\begin{align}
\mathrm{arg}\left(\frac{\mathcal{A}(\overline{D}^0\rightarrow K^- \pi^+ )  }{\mathcal{A}(D^0\rightarrow K^- \pi^+   )} \right) 
	=\mathrm{arg}\left(\frac{  \mathcal{A}(D^0\rightarrow K^+ \pi^-) }{ \mathcal{A}(D^0\rightarrow K^- \pi^+ ) } \right)\,. 
\end{align}

\subsection{Observables for $D^0\rightarrow P^{\pm}V^{\mp}$ decays, $P=K,\pi$, $V=\rho,K^{*}$ \label{sec:complete-system}}

We give the general parametrization of the system of pseudo two-body decays $D^0\rightarrow P^{\pm}V^{\mp}$ in Eqs.~(\ref{eq:uspin-1})--(\ref{eq:uspin-8}) above. We focus here on the lowest-lying $V$ and $P$ states, but analogous expressions hold for (higher) excited states.
The corresponding full three-body decay chains are 
\begin{align}
& \mathcal{A}( D^0\rightarrow \pi^+ K^{*-} \text{ or } K^- \rho^+ \rightarrow K^-\pi^+\pi^0 )\,, \label{eq:3body-1}\\ 
& \mathcal{A}( D^0\rightarrow \pi^{\pm} \rho^{\mp} \rightarrow \pi^+\pi^-\pi^0)\,, \label{eq:3body-2} \\
& \mathcal{A}( D^0\rightarrow  K^{\pm} K^{*\mp}  \rightarrow K^+ K^-\pi^0 )\,,  \label{eq:3body-3} \\ 
& \mathcal{A}( D^0\rightarrow K^+ \rho^- \text{ or } \pi^- K^{*+} \rightarrow  K^+\pi^-\pi^0)\,,  \label{eq:3body-4} \\
& \mathcal{A}( D^0\rightarrow \pi^+ K^{*-} \text{ or } \pi^- K^{*+} \rightarrow K_S\pi^+\pi^- )\,.  \label{eq:3body-5}
\end{align}
Each pair of pseudo two-body amplitudes in Eqs.~(\ref{eq:3body-1})--(\ref{eq:3body-4}) corresponds to seven observables, as in the $D^0\rightarrow \pi^{\pm}\rho^{\mp}$ example (subsection~\ref{subsec:exII}). Four such pairs brings us to 28 observables.
Eq.~(\ref{eq:3body-5}) provides one additional relative phase, making up a total of 29 observables in the general case, which we show in Table~\ref{tab:obsDVP}.

Note that the decays
\begin{align}
& \mathcal{A}( D^0\rightarrow K^+ K^{*-} \rightarrow  K^+K_S\pi^- )\,,\\ 
& \mathcal{A}( D^0\rightarrow K^- K^{*+} \rightarrow  K^-K_S\pi^+ )\,, 
\end{align}
do not generate interference effects with the other $D^0\rightarrow P^{\pm}V^{\mp}$ modes. 

The relations between the phases in Table~\ref{tab:obsDVP} in the CP limit can be understood as follows.
In the CP limit we have, for example
\begin{align}
\mathrm{arg}\left(\frac{\overline{D}^0\rightarrow K^+\rho^-}{D^0\rightarrow K^+\rho^-}\right) 
&= \mathrm{arg}\left(\frac{ D^0\rightarrow K^-\rho^+  }{D^0\rightarrow K^+\rho^-} \right) \label{eq:phases-example-1} 
\end{align}
and
\begin{align}
\mathrm{arg}\left(\frac{\overline{D}^0\rightarrow K^- \rho^+}{D^0\rightarrow K^-\rho^+}\right) 
&= \mathrm{arg}\left(\frac{ D^0\rightarrow K^+\rho^-    }{D^0\rightarrow K^-\rho^+} \right) \label{eq:phases-example-2}\\
&= -\mathrm{arg}\left(\frac{ D^0\rightarrow K^-\rho^+  }{D^0\rightarrow K^+\rho^-} \right) \label{eq:phases-example-3}\\
&= -\mathrm{arg}\left(\frac{\overline{D}^0\rightarrow K^+\rho^-}{D^0\rightarrow K^+\rho^-}\right)\,.\label{eq:phases-example-4}
\end{align}
In Eqs.~(\ref{eq:phases-example-1}) and (\ref{eq:phases-example-2}) we CP-conjugate initial and final state in the numerator.
In Eq.~(\ref{eq:phases-example-3}) numerator and denominator are interchanged, giving a minus sign.
Finally, in Eq.~(\ref{eq:phases-example-4}) we again CP-conjugate initial and final state in the numerator, implying altogether that the LHS of 
Eqs.~(\ref{eq:phases-example-1}) and (\ref{eq:phases-example-2}) are the same up to a minus sign. 

Note that the phase
\begin{align}
\mathrm{arg}\left(\frac{ D^0\rightarrow K^-\rho^+  }{D^0\rightarrow K^+\rho^-} \right) 
\end{align}
only becomes accessible in the SM and CP limit, analogous to the $D\rightarrow PP$ case.

\begin{table}[t]
\begin{scriptsize}
\begin{center}
\begin{tabular}{l|c|c|c}
\hline
\multicolumn{2}{c|}{Observables}  & \multicolumn{1}{p{3cm}|}{Relations in the SM limit, $r_{\mathrm{NP}} = 0$} &\multicolumn{1}{p{3cm}}{Relations in the CP limit, $\mathrm{arg}(\lambda_{b}) = 0 $}   \\\hline\hline
\multirow{8}{3.3cm}{CP-averaged branching ratios} & 
$\,\mathcal{B}(D\rightarrow \pi^+ K^{*-} )$  &  &  \\
& $\,\mathcal{B}(D\rightarrow \pi^+ \rho^- )$  &  &  \\
& $\,\mathcal{B}(D\rightarrow K^+ K^{*-}   )$  &  &  \\
& $\,\mathcal{B}(D\rightarrow K^+ \rho^-   )$  &  &  \\
& $\,\mathcal{B}(D\rightarrow  K^- \rho^+  )$  &  &  \\
& $\,\mathcal{B}(D\rightarrow \pi^- \rho^+ )$  &  & \\
& $\,\mathcal{B}(D\rightarrow K^- K^{*+}   )$  &  &  \\
&$\,\mathcal{B}(D\rightarrow \pi^- K^{*+} )$  &  &  \\\hline
\multirow{8}{3.3cm}{Direct CP asymmetries} &
  $\, a_{CP}^{\mathrm{dir}}(D\rightarrow \pi^+ K^{*-} )$  & $=0$ & $=0$ \\
& $\, a_{CP}^{\mathrm{dir}}(D\rightarrow \pi^+ \rho^- )$  &  &  $=0$ \\
& $\, a_{CP}^{\mathrm{dir}}(D\rightarrow K^+ K^{*-}   )$  &  &  $=0$ \\
& $\, a_{CP}^{\mathrm{dir}}(D\rightarrow K^+ \rho^-   )$  & $=0$ & $=0$ \\
& $\, a_{CP}^{\mathrm{dir}}(D\rightarrow  K^- \rho^+  )$  & $=0$ & $=0$ \\
& $\, a_{CP}^{\mathrm{dir}}(D\rightarrow \pi^- \rho^+ )$  &  &  $=0$\\
& $\, a_{CP}^{\mathrm{dir}}(D\rightarrow K^- K^{*+}   )$  &  &  $=0$ \\
& $\, a_{CP}^{\mathrm{dir}}(D\rightarrow \pi^- K^{*+} )$  & $=0$  & $=0$ \\\hline
\multirow{13}{3.3cm}{Phases from interference effects} &
$\,\arg\left(\frac{\mathcal{A}(\overline{D}^0\rightarrow \pi^+ K^{*-})}{ \mathcal{A}(D^0\rightarrow \pi^+ K^{*-})}\right)$  &    &  \\
& $\,\arg\left(\frac{\mathcal{A}(\overline{D}^0\rightarrow \pi^+ \rho^-)}{\mathcal{A}( D^0\rightarrow \pi^+ \rho^-)} \right)$  &  &  \\
& $\,\arg\left(\frac{\mathcal{A}(\overline{D}^0\rightarrow K^+ K^{*-} ) }{ \mathcal{A}(D^0\rightarrow K^+ K^{*-} ) } \right)$  &  &  \\
& $\,\arg\left(\frac{\overline{D}^0\rightarrow K^+ \rho^-  }{ D^0\rightarrow K^+ \rho^-  } \right) $  &  & \\
& $\,\arg\left(\frac{\mathcal{A}(\overline{D}^0\rightarrow  K^- \rho^+ )}{ \mathcal{A}(D^0\rightarrow  K^- \rho^+ )} \right)$  & $=-\arg\left(\frac{\mathcal{A}(\overline{D}^0\rightarrow K^+ \rho^- ) }{ \mathcal{A}(D^0\rightarrow K^+ \rho^- ) } \right) $  & $=-\arg\left(\frac{\mathcal{A}(\overline{D}^0\rightarrow K^+ \rho^- ) }{ \mathcal{A}(D^0\rightarrow K^+ \rho^- ) } \right) $  \\
& $\,\arg\left(\frac{\mathcal{A}(\overline{D}^0\rightarrow \pi^- \rho^+)}{ \mathcal{A}(D^0\rightarrow \pi^- \rho^+)} \right)$  &  & $=-\arg\left(\frac{\mathcal{A}(\overline{D}^0\rightarrow \pi^+ \rho^-)}{ \mathcal{A}(D^0\rightarrow \pi^+ \rho^-)} \right)$ \\
& $\,\arg\left(\frac{\mathcal{A}(\overline{D}^0\rightarrow K^- K^{*+} ) }{ \mathcal{A}(D^0\rightarrow K^- K^{*+}  )} \right)$  &  & $=-\arg\left(\frac{\mathcal{A}(\overline{D}^0\rightarrow K^+ K^{*-})  }{ \mathcal{A}(D^0\rightarrow K^+ K^{*-})  } \right)$ \\
& $\,\arg\left(\frac{\mathcal{A}(\overline{D}^0\rightarrow \pi^- K^{*+})}{ \mathcal{A}(D^0\rightarrow \pi^- K^{*+})} \right) $  & $=-\arg\left(\frac{\mathcal{A}(\overline{D}^0\rightarrow \pi^+ K^{*-})}{ \mathcal{A}(D^0\rightarrow \pi^+ K^{*-})}\right)$     & $=-\arg\left(\frac{\mathcal{A}(\overline{D}^0\rightarrow \pi^+ K^{*-})}{ \mathcal{A}(D^0\rightarrow \pi^+ K^{*-})}\right)$ \\
& $\,\arg\left(\frac{\mathcal{A}(D^0\rightarrow \pi^+ K^{*-})}{\mathcal{A}(D^0\rightarrow K^-\rho^+)}\right)$   &  &   \\
& $\,\arg\left(\frac{\mathcal{A}(D^0\rightarrow \pi^+\rho^-)}{\mathcal{A}(D^0\rightarrow \pi^-\rho^+)}\right)$  &   &  \\
& $\,\arg\left(\frac{\mathcal{A}(D^0\rightarrow K^+K^{*-})}{\mathcal{A}(D^0\rightarrow K^-K^{*+})}\right)$  &   &   \\
& $\,\arg\left(\frac{\mathcal{A}(D^0\rightarrow K^+\rho^-)}{\mathcal{A}(D^0\rightarrow \pi^- K^{*+})}\right)$  &   &   \\
& $\,\arg\left(\frac{\mathcal{A}(D^0\rightarrow \pi^+ K^{*-})}{\mathcal{A}(D^0\rightarrow \pi^-K^{*+})}\right) $  & $=-\arg\left(\frac{\mathcal{A}(\overline{D}^0\rightarrow \pi^+ K^{*-})}{ \mathcal{A}(D^0\rightarrow \pi^+ K^{*-})}\right)$     & $=-\arg\left(\frac{\mathcal{A}(\overline{D}^0\rightarrow \pi^+ K^{*-})}{ \mathcal{A}(D^0\rightarrow \pi^+ K^{*-})}\right)$  \\\hline
\# of independent observables & \, 29 & \, 22   & \, 16   \\\hline\hline
\end{tabular}
\caption{Available observables in the $D\rightarrow VP$ system, and relations between them in the SM limit and in the CP limit.
\label{tab:obsDVP}}
\end{center}
\end{scriptsize}
\end{table}

\end{appendix}

\clearpage

\bibliography{draft.bib}
\bibliographystyle{apsrev4-1}

\end{document}